\documentclass[12pt]{article}
\usepackage{epsfig}
\usepackage{graphicx}

\def \sect #1 {\setcounter{equation} 0\section{#1}}
\def \be  {\begin{equation}}
\def \ee  {\end{equation}}
\def \ba  {\begin{eqnarray}}
\def \ea  {\end{eqnarray}}
\def \baa {\begin{eqnarray*}}
\def \eaa {\end{eqnarray*}}
\def \bb  {\begin{thebibliography}}

\def \lab #1 {\label{#1}}

\def \e {\mbox{e}}

\def \fracs #1#2 {\mbox{\small $\frac{#1}{#2}$}}

\def \bin #1#2 {{\left({#1}\atop{#2}\right)}}
\def \as {\relax\ifmmode\alpha_s\else{$\alpha_s${ }}\fi}

\def \al #1 {\frac {\as({#1})}{\pi} }
\def \ds #1 {\ooalign{$\hfil/\hfil$\crcr$#1$}}

\newcommand \bea{\begin{eqnarray}}
\newcommand \eea{\end{eqnarray}}

\newcommand \ep {\epsilon}

\def \o {\omega}
\def \ve {\varepsilon}
\def \g {\Gamma}
\def \a {\alpha}
\def \p {\pi}
\def \r {\rho}
\def \m {\mu}
\def \n {\nu}
\def \ep {\epsilon}

\def \la {\lambda}
\def \s {\sigma}
\def \x {\xi}
\def \f {\phi}

\def\hepph  #1 {{\tt hep-ph/#1}}

\begin{document}

\begin{flushright}
YITP-SB-04-59
\end{flushright}

\vbox{\vskip .5 in}

\begin{center}
{\Large \bf QCD and Jets\footnote{
To appear in proceedings of the
2004 Theoretical Advanced Study Institute in Elementary
Particle Physics (TASI 2004), University of Colorado at Boulder,
June 6 - July 2, 2004.}
}

\vbox{\vskip 0.25 in}

{\large George Sterman}

\vbox{\vskip 0.25 in}

{\it C.N.\ Yang Institute for Theoretical Physics,
Stony Brook University, SUNY\\
Stony Brook, New York 11794-3840, U.S.A.}
\end{center}

\begin{abstract}
These lectures introduce some of the basic methods
of perturbative QCD and their applications to
phenomenology at high energy.
Emphasis is given to
techniques that are used to study
QCD and related field theories to all orders in perturbation theory,
   with introductions to infrared safety, factorization and evolution in
high
energy hard scattering.
\end{abstract}

\bigskip


\section{Introduction}

Quantum chromodynamics is the component of the Standard
Model that describes the strong interactions.  It is a
theory formulated in terms of quarks and gluons at the Lagrangian
level, but observed in terms of nucleons and mesons in nature.
Partly because of this, its verification required formulating new
methods and asking new questions on how to confront quantum
field theories with experiment, especially those that exhibit strong
coupling.
In what might seem an almost paradoxical
outcome, new roles were recognized for perturbative methods
based on the elementary constituents of the theory.

Since strong coupling is a feature of many extensions of the
Standard Model and of string theories,
these theoretical developments in
perturbative QCD remain of interest in a more general context.
In addition, the perturbative analysis of hadronic scattering and
final states played an essential role in the verification of the
electroweak
sector of the Standard Model, through the discoveries of the
W and Z bosons and of the bottom
and top quarks.  Likewise, perturbative methods are indispensable
in the search for new physics at
all colliders, which will keep them relevant to phenomenology
for the foreseeable future.
Finally, at lower but still substantial energies, new
capabilities to study the polarization-dependence of hadronic
scattering and
the strong interactions at high temperatures and densities
are opening new avenues for the study of quantum field theory.

These lectures, of course, cannot cover the full range of QCD,
even at high energies in perturbation theory.
They are meant to be an introduction for students
familiar with the basics of quantum field theory
and the Standard Model, but not with applications of perturbative QCD.
I will emphasize
the  formulation and calculation of observables using perturbative
methods, and discuss the justification of perturbative calculations
in a theory like QCD, with esssentially nonperturbative long-distance
behavior.
I will also introduce elements of the phenomenology of perturbative QCD
at high energies, keeping in mind its use in the search
for signals of new physics.

\section{QCD from Asymptotic Freedom to Infrared Safety}

We begin with a quick prehistory of quantum
chromodynamics, and of the strands of experiment
and theory that converged in the concept of asymptotic freedom.
We also show how infrared safety makes possible
a wide range of phenomenological implementations
of asymptotic freedom.

\subsection{Why QCD?}

The search that culminated in the $SU(3)$ gauge theory
quantum chromodynamics was triggered by the discovery of
an ever-growing list of hadronic states, including mesons, excited
nucleons and a variety of resonances, beginning with the
$\Delta$ resonances first seen in the early 1950s \cite{pais},
and continuing to this day \cite{pdg}.   Already in 1949, the idea
was put forward that mesons might be composites
of more elementary states (of nucleons in the very early work of
Fermi and Yang \cite{fermiyang}).
Clearly  hadrons were not
themselves elementary in the normal sense,
but for a long time the
strong couplings between hadronic states, and their
ever-growing number, left the situation unsettled.

   In broad terms, two alternative viewpoints
   were entertained  for nucleon and other hadronic states.
In bootstrap scenarios, each
hadronic state was at once deducible from the others,
yet irreducible, in the sense
that no hadronic state was more elementary than any other.
This idea eventually evolved through the dual model
\cite{dualmodel} into string theory, leaving
the strong interactions behind, at least temporarily.
The more  conservative viewpoint regarded
hadrons as composite states of elementary particles, but
the observed couplings seemed to pose an
insurmountable barrier to a field-theoretic treatment.
Still, the growing success of quantum electrodynamics
held out at least the possibility that such a picture might
somehow emerge.

With the weak and electromagnetic interactions
serving as diagnostics, what we now call flavor symmetry
was developed out of the
spectroscopy of hadronic states.
Their interactions through local currents also
hinted at point-like substructure.  The weak and electromagnetic
currents,
in their turn, were naturally described
as coupling to  fundamental objects, and the quark model was born,
partly as a wonderfully compact classification scheme,
and partly as a dynamical model of hadrons.

Increases in accelerator energies
eventually clarified the situation.
A prologue, both scientifically and technologically,
   were the 1950's measurements of
   electron-proton elastic scattering \cite{epelastic}.
Imposing the relevant symmetries of the
electromagnetic and strong interactions
(parity and time-reversal invariance),
the cross section for electron-nucleon unpolarized scattering at
   fixed angle is customarily expressed in the nucleon rest frame,
   as
   \be
{{d\sigma}\over{d\Omega_e}}\ =\
\left[\, {\alpha_{\rm EM}^2 \cos^2(\theta/2) \over 4 E^2
\sin^4(\theta/2)}\, \right]\,
{E' \over E}\, \left( {{| G_E(Q) |^2\ +\ \tau | G_M(Q) |^2}\over
{1+\tau}} \ +
\  2\tau | G_M(Q)|^2 \tan^2{\theta /2}  \right)   .
\label{Mottsigma}
\ee
The term in square brackets is the (Mott) cross section for the
scattering of
a relativistic electron of energy $E$ by a fixed
source of unit charge, while $E'$ is the outgoing electron energy.
The variable
$\tau\ \equiv \ Q^2/4M_N^2$, with $M_N$ the nucleon mass
and $Q^2=-EE'\sin^2(\theta/2)$
the invariant momentum transfer.  The
functions $G_E(Q)$ and $G_M(Q)$ are the electric and magnetic
form factors.  They fall off rapidly with momentum transfer, roughly as
\bea
|G(Q)_i| \sim {1\over (1 +  Q^2/\mu_0^2)^2}\, ,
\eea
where $\mu_0\sim 0.7 GeV$.
This simple behavior is anaolgous to
the form factors of nonrelativistic bound states
with extended charge distributions \cite{perl}.
This suggests a substructure
for the nucleon,  but leaves its nature
mysterious.  Indeed, much
is still being learned about nucleon
form factors \cite{jlabform} and their
relation to the quarks and gluons of QCD \cite{qcdform,dvcs}.  The
experiments
are difficult, however, because of the form
factors'  rapid decrease with $Q$.  This is simply to say  that when
a proton is hit hard, it is likely to radiate
or to break apart.

As available electron energies increased it became
possible by the late 1960's to go far beyond
elastic scattering, and to study highly (``deep")
inelastic reactions.  In these experiments, nature's choice
became manifest.

\begin{figure}[h]
\hbox{\hskip 2 in {\epsfxsize=8cm \epsfysize=8cm \epsffile{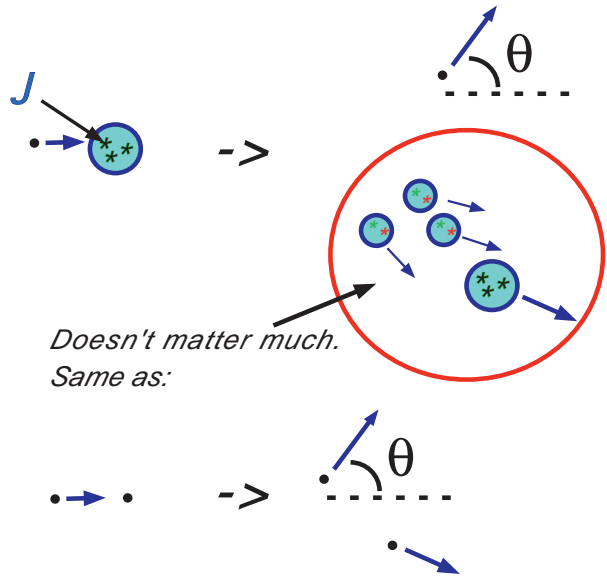}}}
\caption{Picture of deep-inelastic scattering}
\label{disfig}
\end{figure}

The idea is simple, and analogous
in many ways to the famous Rutherford experiments
that revealed the atomic nucleus.  A high energy
electron beam is made to collide with a nucleon target.
One observes the recoil of the electron,
measuring the total cross section as a function of
transfered energy and scattering angle, as indicated in
Fig.\ \ref{disfig}.  Although each individual event is
generally quite complicated, a simpler
pattern emerges when events at a given
momentum transfer are grouped together.  Imposing
again the relevant symmetries,
a general deep-inelastic scattering (DIS) cross section can be written
in terms of dimensionless structure functions, $F_1(x,Q^2)$
and $F_2(x,Q^2)$, where the variable $x$ is defined as
\bea
x = {Q^2 \over 2 p_N\cdot q}\, ,
\label{xdef}
\eea
with $q$, $q^2=-Q^2$, the momentum transfer as above.
In these terms, the  cross section (again in the nucleon rest frame) is
analogous to the one for elastic scattering,
\be
{d\sigma \over dE' \, d\Omega}
=
\left[\, {\alpha_{\rm EM}^2 \over 2SE\, \sin^4(\theta/2)}\, \right]\;
\left( 2\sin^2(\theta/2) F_1(x,Q^2) + \cos^2(\theta/2)\,
\frac{m}{E-E'}\, F_2(x,Q^2) \right)\, ,
\label{sigmaasFs}
\ee
   where $S$ is the overall center of mass energy squared.
   The factor in square brackets is again proportional to the
   cross section for scattering from a point charge by
   single-photon exchange.  The complexities due to the
hadron target are all in the structure functions.
   The structure functions, however, turn out to
   be simpler than might have been expected.
    To a good approximation in the original experiments,
   the $F$'s were found to be  indpendent of momentum transfer.
   This surprising result, called scaling, was in contrast to the rapid
decrease
   of elastic form factors.

   It is as if, for the purposes of
   inclusive scattering, the $Q$-dependence is given
   entirely by the elastic scattering of an electron from a
   point charge.
   Representing these point charges by $\star$ as in the figure,
    we can write schematically,
   \be
\sigma_{\rm eN\to e X}(x,Q)\ \sim
\sigma^{\rm (elastic)}_{\rm e\star\to e\star}(x,Q)\, \times\, {{\cal
F}_N(x)}\, ,
\label{dissig}
\ee
where ${\cal F}_N$ is a linear
combination of the structure functions and where
$\sigma^{\rm (elastic)}_{\rm e\star\to e\star}(x,Q)$ is the
lowest-order cross section for the scattering of an electron
on an elementary $\star$, with $x$ given by (\ref{xdef}).  The precise
cross section depends on the
spin of the $\star$'s, which are identified as the charged {\it partons}
inside the proton \cite{feynpm}.   The variable $x$ can then be given a
nice physical interpretation, if we think of the
scattering in a frame where the nucleon and all its partons
are moving in the  same direction at velocities approaching the speed
of light,
the eN center of mass frame for
example.
In this frame, if the nucleon
momentum is $p_N$, the parton's momentum should be some fraction
of that, call it $\xi p_N$, with $0<\xi<1$.
   Then if the scattered parton $\star$ is to be
on-shell, and and if we neglect its mass in the kinematics
of the hard scattering, we find
\be
(\xi p_N+q)^2=0 \quad \rightarrow \quad \xi ={Q^2\over 2p_N\cdot q}=x\,
.
\ee
Thus the scaling variable $x$ observed in a specific event is
the momentum fraction of the parton from which the electron
has scattered.

We can now interpret the $x$-dependent structure functions for DIS as
probabilities to find partons with definite momentum fractions,
\bea
\sigma_{\rm eN\to e X}(x,Q)\ \sim
\sigma^{\rm (elastic)}_{\rm e\star\to e\star}(x,Q)\ \times\ {\rm
(probability\ to\ a\ find\ parton)}\, .
\eea
The precise form of the parton-electron cross section
depends on the spin of the parton through the structure functions.
In this way, early experiments could already
identify the spin as 1/2, and the partons
quickly became identified with quarks.

The parton model gave an appealing physical picture
of the early deep-inelastic scattering experiments, but it seemed
to contradict other experimental facts.
If the cross section can be computed with lowest-order electromagnetic
scattering, then partons must in some sense be weakly bound.
But then what happened to the strong force?   Already at
that time searches for free quarks had come up empty, and
the concept of quark confinement was taken as yet another
sign of the strength of nucleon's binding.  How was it possible
to describe such a force, strong enough to confine yet
weak enough for scaling?

\subsection{Quantum chromodynamics and asymptotic freedom}

Scaling posed the question to which QCD was the answer.  The essential
feature
of QCD is
asymptotic freedom \cite{asyfree}, according to which its coupling,
$g_s$ decreases
toward short distances and increases toward longer distances and times.
Asymptotic freedom matches qualitatively with the requirements of
approximate scaling,
and its converse, that the coupling increases toward longer distances
and times, is consistent with the behavior that gave the strong force
its name.
As we shall see, asymptotic freedom also provides systematic and
quantitative predictions
for corrections to scaling.  Let's recall first how the strength of a
coupling
varies with distance scale.

The (classical) Lagrangian of QCD is \cite{fgmlqcd}
\be
{\cal L} = \sum_{f=1}^{n_f} \bar q_f\, \left(
i\rlap{/}\partial-g_s\rlap{/}A +m_f\, \right)q_f -
{1\over 2}\,  {\rm Tr} (F_{\mu\nu}^2[A])\ ,
\ee
in terms of $n_f$ flavors of quark fields and $SU(3)$ gluon fields
$A^\mu = \sum_{a=1}^8A^\mu_a T_a$, with $T_a$ the generators
of $SU(3)$ in the fundamental representation.  The field strengths
$F_{\mu\nu}[A]=\partial_\mu A_{\nu}-\partial_\nu A_{\mu}
+ ig_s[A_\mu,A_\nu]$ specify the familiar three- and four-point
gluon couplings of nonabelian gauge theory \cite{yangmills}.  We will
not need their explicit forms here.
This Lagrangian, with its exact color gauge symmetries, also
   inherited  the approximate flavor symmetries
of the quark models, and was automatically consistent with
properties and consequences of hadronic weak interactions that follow
from an analysis of vector and axial vector currents \cite{weiqcd}.
In addition to the right currents, QCD had just the right kind of
forces.

\begin{figure}[h]
\centerline{{\epsfxsize=7.5cm  \epsffile{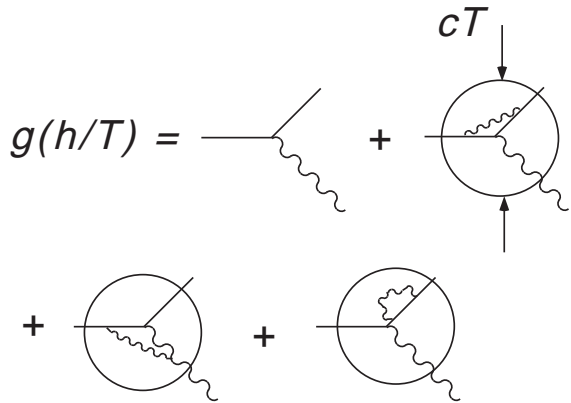}}}
\caption{The coupling normalized in position space}
\label{couplingfig}
\end{figure}

A simple picture (if not particularly convenient for
calculation) of the scale dependence of the  coupling is shown in
Fig.\ \ref{couplingfig}, which represents the perturbative series for
the
amplitude for fermion-vector coupling.  We imagine computing this
quantity in coordinate space, which amounts
to integrating over the positions of vertices.  At lowest order, this
amplitude is just
the coupling ($e$ in QED, $g_s$ for QCD, etc.).  At the one-loop level,
the amplitude
requires renormalization because whenever two vertices,
say at $x$ and $y$,
coincide, a propagator connecting them diverges.
For example, the coordinate space propagator
for a  massless boson is
\bea
\frac{1}{(2\pi)^2}\, {1 \over (x-y)^2-i\ve} = i\, \int {d^4k \over
(2\pi)^4}\, {\rm e}^{-ik\cdot (x-y)} \;
{1\over k^2 +i\epsilon}\, .
\label{propxy}
\eea
Alternatively, we can just notice that the integrals over the positions
of the vertices are dimensionless in gauge theories.
Such integrals have no choice but
to diverge for ultraviolet momenta, where masses are negligible.

In these terms, the  renormalized coupling may be defined (in Euclidean
space)
as the sum of all diagrams with vertices contained within  a
four-dimensional
sphere of diameter $cT$, with $c$ the speed of light and $T$ any fixed
time.
Let's call the coupling defined this way $g_s(h/T)$, where $h/T$ is
the corresponding energy scale, which is then the renormalization scale.
Although we cannot compute $g_s(h/T)$ directly,
we can compute the  right-hand side of the equation
\bea
T\, {d\over dT}\ g_s(h/T) =  \frac{b_0}{16\pi^2}\, g_s^3(h/T) +
\cdots\, ,
\label{run}
\eea
where  the term shown on the right is the finite result found
when one of the vertices is fixed at the edge of the sphere.
(We shall neglect higher orders, assuming that $g_s$ is  small.)
In effect, the divergent part of the integrals is independent of $T$.
The constant $b_0$, of course, depends on the  theory.  For
QCD, we famously have $b_0=11-2n_f/3$, with $n_f$ the number  of quark
flavors.  $g_s(h/T)$ therefore decreases as $T$ decreases, and
QCD is asymptotically free.
The mechanism of asymptotic freedom in gauge theories
is often compared to the antiscreening, or strengthening, of applied
magnetic fields
in paramagnetic materials (whose internal magnetic moments line up
with the external field).  This behavior, associated with the
self-coupling
of the gluons, gives rise to the 11 term in $b_0$.  The $-2n_f/3$ term
is the competing influence of $n_f$ flavors of quarks,
whose effect is to screen color, in analogy to screening
in diamagnetic materials (whose internal moments oppose an applied
field).

The solution to Eq.\ (\ref{run}), expressed in terms of
the QCD analog of the fine-structure constant,
$\as(\mu)\equiv g_s^2(\mu)/4\pi$  is central to the interpretation of
QCD,
\bea
\as(\mu') &=& \frac{\as(\mu)}{1+b_0{\as(\mu)\over 4\pi}\,
\ln(\mu'{}^2/\mu^2)}
\nonumber\\
&\equiv& {4\pi\over b_0\, \ln\left[{\mu'\over \Lambda_{\rm
QCD}}\right]^2}\, .
\label{alphamumuprime}
\eea
In the first expression we relate the coupling defined by time
$T=h/\mu$ to the coupling defined by
another time $T'=h/\mu'$.  In the second expression, we
use the observation that in nature it does not matter which
starting scale $\mu$ we choose; all must give the same
answer for $\as(\mu')$. The scale $\Lambda_{\rm QCD}=\mu
\exp[2\pi/b_0\as(\mu)]$
is an invariant, independent of $\mu$, and serves to set consistent
boundary
conditions for the solution of (\ref{run}).

A simple but important observation is
that although the coupling $g_s(\mu)$ changes with the renormalization
scale $\mu$,
no physical quantity, $\Pi(Q)$ depends on $\mu$, where $Q$ stands for
any external momentum or physical mass scale(s).  On the other hand,
assuming that
we can compute $\Pi(Q)$ in
perturbation theory it will be an expansion in $\alpha_s(\mu)$,
and the renormalization scale will also appear in ratios $Q/\mu$
with physical scales.
For such a perturbative expansion, we have the consistency condition
\be
\mu{d\Pi\left(Q,Q/\mu,\alpha_s(\mu)\right)\over d\mu}=0\, .
\label{physzero}
\ee
We will come back to this relation several times in what follows.

The qualitative picture for DIS in Fig.\ \ref{disfig} now has a
natural explanation at large energy and momentum transfer.
During the short time that it takes the electron to exchange
a virtual photon, the strong force acts as though it were weak,
and the electromagnetic scattering of a quark proceeds as
though the strong force were almost irrelevant.  The scattered quark,
however, starts to move away from the target, and over
a time scale of the order of the nucleon size, the strong force
acts strongly, and produces the details of the inelastic
final state, as the disturbed system of partons
reassembles into hadrons.  By this time, however, the electron is long
gone,
and the distributions we measure are detemined primarily by
the original, electromagnetic scattering, which measures,
as indicated in Eq.\ (\ref{dissig}), the probability of finding quarks
of various momenta in the parton.  As the process just
described proceeds, the value of relevant coupling changes
dramatically in magnitude, from small to large.

The discovery of asymptotic freedom opened a new chapter,
not only in the strong interactions, but in relativistic quantum field
theory.
For the first time, effects at all orders in perturbation theory
gave rise to observable consequences, and even the qualitative
features of experiments could not be understood without them.

The explanation of scaling by
asymptotic freedom played the same role for hadronic physics
that the explanation of elliptical orbits
from the law of gravitation played
for celestial mechanics.
In both cases, most subsequent applications
were (and are) to much more complicated experiments (scale breaking,
hadron-hadron scattering for QCD, the many-body problem
for gravitation).   The striking initial success led
to an essentially open-ended process of learning how to
use the new theory.
For QCD, the problem was, and still is,  to calculate
in a theory that includes strong as well as weak coupling,
and whose short distance, fundmental degrees of freedom (quarks and
gluons)
do not appear as asymptotic states (where we see only hadrons).

\subsection{Learning to calculate in the new theory}

One ultimate goal of studies of quantum chromodynamics might be
to explain nuclear physics, say at the level that quantum
electrodynamics explains chemistry.  But when QCD was new
it was natural to wonder whether
we could even study the elementary quanta of the  theory, the quarks
out of which we construct the electroweak currents, and the gluons that
give
rise to the forces.  As a practical matter, these particles are confined
in bound states characterized by strong forces.
Although the parton model affords a lively and compelling
picture of a specific process, deep-inelastic scattering, it was
by no means obvious that DIS was not a special case, or
that QCD had predictions that would be realized in
other experiments.

The total cross section for deep-inelastic scattering is special
because it can be related directly to the expectation value
of a product of operators at short distances.  It works this way.
DIS is initiated by the exchange of a virtual photon emitted by
an electron.  The photon
couples to quarks via the electromagnetic current, $J_\mu(x)$, and
the amplitude to produce some hadronic final state $|X\rangle$
from nucleon state $|p_N\rangle$ is governed by the matrix elements
$\langle X|J_\mu(0)|p_N\rangle$.  In these
terms, the unpolarized, inclusive deep-inelastic
cross section  at momentum transfer $q$ is determined by
the squared hadronic matrix elements (here simplified to a trace)
\bea
\sum_X \left|\, \langle X|J_\mu(0)|p_N\rangle\, \right|^2 \;
\delta^4(p_X-p_N-q)
= \int {d^4y \over (2\pi)^4} \; {\rm e}^{-iq\cdot y}
\langle p_N\, |\, J^\mu(y)\, J_\mu(0)\, |p_N\rangle\, .
\label{inclusive}
\eea
Scaling can be understood as a property of these matrix elements
\cite{bjorken69}.
The large $Q^2=-q^2$ behavior of the Fourier transform
to momentum space is
related to the short-distance behavior of the expectation
value of the product of
currents \footnote{More specifically,
it is the small-$y^2$ or ``light-cone" behavior.}.
These expectation values are observables, and we can therefore
use Eq.\ (\ref{physzero}) above to derive
\bea
\langle p_N\, | J^\mu(y)\, J_\mu(0)\, |p_N \rangle &=& \left
(y^2\right)^{-2}\,
C\left(  {y^2\over \mu^2},p_N\cdot y, \alpha_s(\mu^2)\right)
\nonumber\\
&=& \left (y^2\right)^{-2}\, C\left( 1, p_N\cdot y,
\alpha_s(1/y^2)\right)
\nonumber\\
&=& \left (y^2\right)^{-2}\, \sum_n c_n(\as(1/y^2))\, (p_N\cdot y)^n\, ,
\label{JJlcexp}
\eea
where we have exhibited the overall dimensional behavior as a factor of
$y^2$
and where in the last line we have expanded in powers of the
dimensionless
variable $p_N\cdot y$, a procedure known as the
light-cone expansion \cite{lce,chm}.
After the Fourier transform, the expansion in $p_N\cdot y$
   generates dependence
on the scaling variable, $x$ as in the structure functions of
Eq.\ (\ref{dissig}).
The coefficients $c_n$, however, have the  interpretation of
renormalized expectation values of local operators.  The $y^2$, and
hence $q$-dependence
remains calculable.   This is an exceptional case, however, and
this analysis depends
on our ability to sum over states in Eq.\ (\ref{inclusive}).

The perturbation theory-friendly structure of the fully inclusive cross
section is
to be contrasted to a generic S-matrix element, which depends
on a variety of kinematic variables $Q_i$ and mass scales $m_i$.  It
will not generally
be possible to pick a single renormalization scale to absorb
the dynamical dependence of more than one such scale, and
perturbation theory has little or nothing to say about most S-matrix
elements
in QCD.

For some time after the discovery of asymptotic freedom, it was unclear
whether its use would be limited to those relatively few,
fully inclusive processes that could
be reduced to expectation values of products of currents, or whether
perturbation theory could
ever say anything useful about the structure of final states.
Correspondingly,
it was unclear whether the quarks and gluons of QCD would manifest
themselves in experiment at all.  The search for footprints of partons
took
at least part of its inspiration indirectly, from an unexpected source.

\subsection{The structure of final states:  cosmic rays to quark pairs}

In the mid 1950s, not long after the time when nonabelian gauge theories
were first developed by Yang and Mills \cite{yangmills}, jets began to
be observed
in cosmic ray events \cite{cosmicray}.  The term was applied to sprays
of particles
emerging from the collisions of energetic cosmic rays in an emulsion
target.  The particles typically had large (multi-GeV) and nearly
parallel
longitudinal momenta, with relatively small transverse momenta,
typically of order 0.5 GeV.    Such events are still difficult to
describe fully.  In the context of the parton
model, however, their very structure suggests that, even if
individual quarks might not be observable,  they might still
give rise to jet-like configurations of hadrons.

\begin{figure}[h]
\hbox{\hskip 2 in {\epsfxsize=10cm  \epsffile{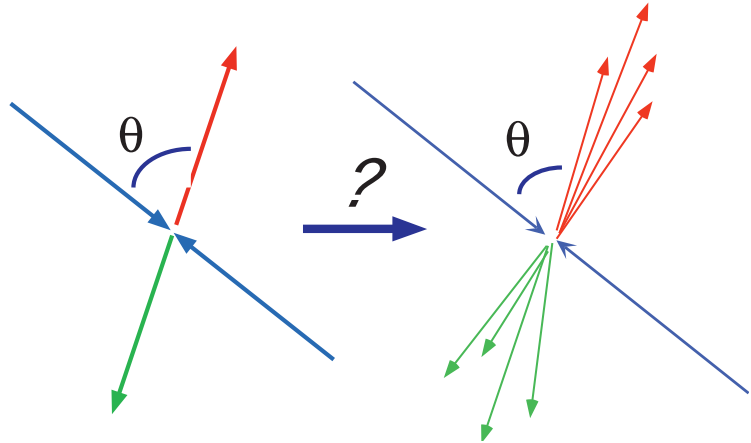}}}
\caption{The parton to hadron transition for two jet events.}
\label{pm2jet}
\end{figure}

The development of a GeV-range electron-positron collider at SLAC
made it possible to
test this idea directly.
In the quark-parton model, even without QCD, a pair of quarks
$q\, \bar{q}$ could be produced by the  annihilation of
an electron-positron pair through a virtual photon,
\be
e^+ e^- \rightarrow \gamma^*(Q) \rightarrow q + \bar{q}\, .
\label{epem}
\ee
Now the total cross section for $\rm e^+e^-$ annihilation
into hadrons can be put into a form very much like Eq.\
(\ref{inclusive})
for deep-inelastic scattering, in terms of the products of currents.
As a result, the total cross section is one of those special
quantities for which perturbation theory can be applied directly.
On the other hand, the lowest-order diagram for the
processs (\ref{epem}) gives  the prediction that the
quarks will emerge from the scattering with a $1+\cos^2\theta$
angular distribution relative to the  beam axes in the center
of mass frame, with $\theta$ the angle between the incoming electron
and the outgoing quark.  If the quark and  antiquark each
give rise to a jet of particles in their respective directions,
the structure of final states will reflect directly the  underlying
partonic process.  To represent the process in the parton model it was
necessary to
implement a model of hadron production within the jet that included
a cutoff in transverse momentum \cite{drellyanjet}.  The picture of the
transition from
partons to hadrons is represented by Fig.\ \ref{pm2jet}.

\begin{figure}[h]
\hbox{\hskip 3 cm {\epsfig{file=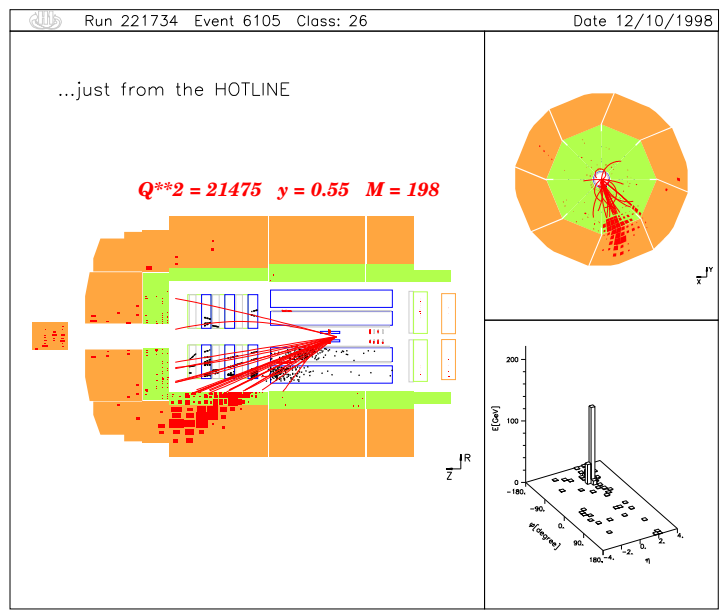,width=80mm,angle=90,clip=}}}
\caption{A current jet in DIS.  From the H1 Collaboration.}
\label{currentjet}
\end{figure}

\begin{figure}[h]
   \hbox{\hskip 5 cm \epsfxsize=8cm \epsffile{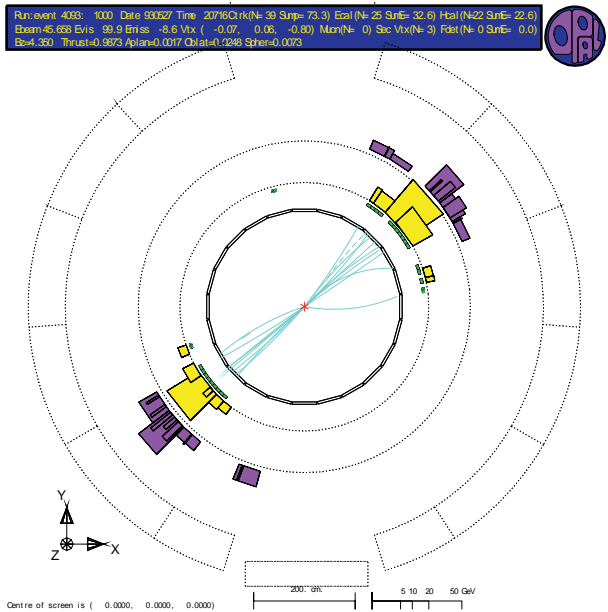}}
\caption{Two jet event in $\rm e^+e^-$ annihilation. From the OPAL
Collaboration.}
   \label{lep2jet}
\end{figure}

Would this picture be realized?  It was, once the center of mass energy
reached about 7 GeV
\cite{han75}, although to see it  at these
energies took some analysis.  Energies have increased
a hundred-fold in the intervening time, and jets have become
a commonplace in high energy physics, in the final states of $\rm
e^+e^-$
annnihilation, of deep-inelastic scattering, and of hadron-hadron
scattering, as illustrated by the next three figures.  In deep-inelastic
scattering, for example, Fig.\ \ref{currentjet}, the so-called
``current jet" associated with
the scattered quark is clearly visible.  Fig.\ \ref{lep2jet} shows
how well-defined two-jet events can be in electron-positron
annihilation at energies around one hundred GeV, and
Fig.\ 6 shows a nice example of jets
from proton-antiproton scattering in the TeV energy range.

    \begin{figure}[h]
    \hbox{\hskip 4 cm\epsfxsize=7.5cm \epsffile{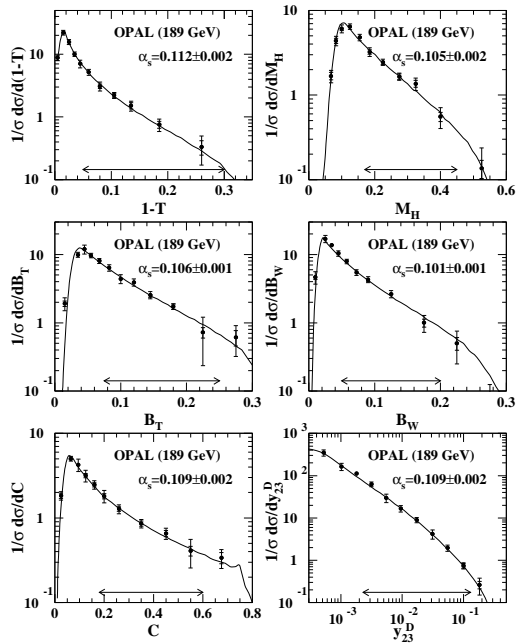}}
    \label{legojets}
    \caption{A ``lego plot" representation of dijets in
    proton-antiproton  scattering,
    represented in the space of rapidity and azimuthal angle with
    respect to the beam axis.  From the D0 Collaboration.}
    \end{figure}

\subsection{Jets in QCD}

By the time jets were first observed, asymptotic freedom had
been discovered, and the basics of the analysis of DIS, which
we will review below, had been developed.
The parton model had predicted the angular distributions for jets
that are initiated by electroweak processes such as $\rm e^+e^-$
annihilation and deep-inelastic scattering, but gave little
guidance on what those jets would actually look like, although it
was thought that, as in cosmic ray events,
transverse momenta might be limited on the scale
of some fraction of a GeV.   At first, it was unclear what QCD
would have to say on this matter, until it was realized that
perturbative field theory can make well-defined predictions
for certain observable quantities even though they
cannot be written in terms of correlation functions
at short distances.  These observables are not fully
inclusive in the hadronic final state, but they do
sum over many final states.  They are
called ``infrared safe" \cite{ste77,irsafe}. Suitably-defined jet cross
sections
are part of this class.
To motivate infrared safety in QCD, we briefly recall the treatment of
infrared
divergences in quantum electrodynamics.

In QED, infrared divergences
are a direct result of the photon's zero mass.  One-loop corrections to
QED processes are typically finite if the photon is given a small mass
$\lambda$,
but diverge logarithmically  as $\lambda$ is taken to zero, through
terms like
\be
\Delta\sigma_{A\to B}^{(1)}\,
\left(Q,m_e,{m_\gamma=\lambda},\alpha_{\rm EM}\right)
\sim \alpha_{\rm EM}\; \sigma_{A\to B}^{(0)}(Q,m_e)\,
\beta_{AB}(Q/m_e)\; {\ln {\lambda\over Q}}\, .
\ee
In this expression, $\sigma_{A\to B}^{(0)}$ is a lowest order cross
section,
characterized by momentum transfer $Q$
and $\beta_{AB}$ is a function that is independent of $\lambda$.
The well-known solution of this problem \cite{blochn} is to introduce
an energy resolution, $\Delta E \ll Q$, and to calculate
cross sections that sum over the emission of all soft photons
with energies up to $\Delta E$.  This is a realistic procedure, since
any measuring
apparatus has a minimum energy sensitivity, below which
soft photons will be missed in any case.  The outcome of
this procedure is that corrections to cross sections including an energy
resolution take the form,
\be
\overline\Delta{\sigma}_{AB}\left(Q,m_e,\Delta E,\alpha_{\rm
EM}\right)
\sim \alpha_{\rm EM}\; \sigma_{A\to B}^{(0)}(Q,m_e)\,
\beta_{AB}(Q/m_e)\; {\ln {\Delta E\over Q}}\, .
\label{enressig}
\ee
The logarithm of the photon mass has been replaced by a logarithm
of the energy resolution.  Notice that if the ratio $\Delta E/Q$ is not
very very small,
and if the function $\beta_{AB}$ is not very large,
   $\alpha_{\rm EM}\sim 1/137$ ensures that (\ref{enressig}) is
numerically small.
This explains the success of many lowest-order  calculations in QED,
in the face of that theory's infrared divergences.

The Bloch-Nordseick procedure in QED suggests that something
similar might be possible in QCD.  In this case, however, since
we want to use perturbation theory at high energy, we must be able to
calculate in
the limit where the ratios of light quark masses to the
energy scale become vanishingly small, along with those involving the
   gluon mass, which is zero to start with.   This leads to a whole new
set of divergences,
   called collinear divergences, which we will have occasion to
   discuss at length below.  Something similar happens if we
   take the electron mass to zero in Eq.\ (\ref{enressig}), in which case
the functions $\beta_{\rm AB}$ develop logarithmic
singularities.

Neglecting for the moment heavy quarks, we are
after a set of cross sections that are perturbatively finite even
when all masses in the theory are set to zero.  To go to high
energy in perturbation theory, we must thus check that we can go to
the zero-mass limit.  Once we can identify a quantity with a finite
zero-mass limit, and have traded zero mass back
for high energy, we have a situation that is perfect for QCD.
We will be able to use (\ref{physzero}) to pick the coupling
at the scale of the energy, and asymptotic freedom
will ensure that as the energy scale grows,
the relevant coupling will decrease.  Perturbative predictions will then
improve with increasing energy.

The classic analyses of Kinoshita and of Lee and Nauenberg \cite{kln}
showed that
{\it total} transition rates remain finite in fully massless theories
because the zero-mass limit does not violate unitarity in perturbation
theory.
Infrared safe cross sections are generalizations of this
analysis to less inclusive observables.  For QED, this
can be done with an energy resolution; for QCD in the zero-mass limit,
this is not sufficient.   For
$\rm e^+e^-$ annihilation, however, we can identify infrared safe
quantities
by introducing an additional resolution.  The motivation is
completely analogous to the QED case.  In the limit of zero quark
mass, a quark of momentum $p,\, p^2=0$ can emit an gluon of
momentum $xp,\, 0<x<1,\, (xp)^2=0$ and remain on-shell, since
the remaining momentum $(1-x)p$ is still lightlike with positive
energy.  The resulting quark and
gluon, however, are exactly collinear in direction, and it is by no
means
clear how to resolve them, especially since the emission, or
its inverse, can take
place at any time, even within a hypothetical detector.  The same
would be true for a massless electron and collinear photon.

\begin{figure}[h]
\centerline{{\epsfxsize=7cm  \epsffile{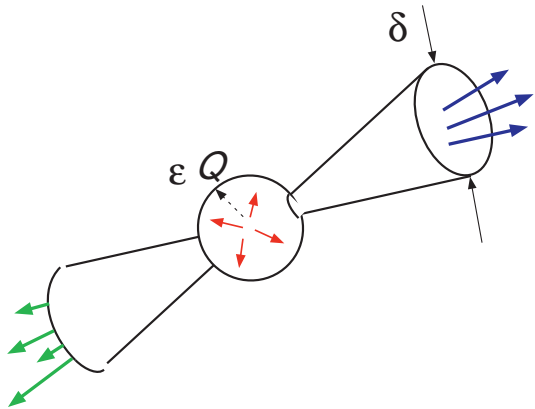}}}
\caption{Cone jets for $\rm e^+e^-$ annihilation.}
\label{conejet}
\end{figure}

If we draw an analogy to the energy resolution of  QED,
we are naturally led to
seek observables with angular as well as energy resolutions
   for high energy QCD (or massless QED),
as represented in Fig.\ \ref{conejet}, where the cones show
an angular range into which large energy flows, while the small
ball in the remaining directions represents an energy resolution.
Without going into detail yet, such cross sections are infrared safe,
and depend only on the overall energy $Q$, the angular resolution
$\delta$,
and the energy resolution $\epsilon Q$, with $\epsilon$ a small but
finite number.
Because they are physical quantities,
   the perturbative expansions for the
   corresponding cross sections satisfy Eq.\ (\ref{physzero}), and we can
write
   \bea
   \sigma_{\rm jet}\left(Q/\mu,\delta,\epsilon,\alpha_s(\mu)\right)
&=&
\sigma_{\rm jet}\left(1,\delta,\epsilon,\alpha_s(Q)\right)
   \nonumber\\
   &=& \sum_{n\ge 0} \Sigma_n(\delta,\epsilon)\, \alpha_s^n(Q)\, .
   \eea
We can thus calculate these cross sections directly in perturbation
theory \cite{ste77}, without needing to impose cutoffs in transverse
momentum.  Infrared safety alone is sufficient to produce jet
structure at high energies, since each emission outside the
cone is suppressed by a factor of $\as(Q)$.
The two-jet cross sections of $\rm e^+e^-$ of Fig.\ \ref{conejet}
can be generalized in a number of ways that we will discuss below.

We may formalize the concept of infrared
safety in the following way.
QCD perturbation theory gives self-consistent predictions
for a quantity $C$ when $C$:

   \begin{itemize}

\item {is dominated by short-distance dynamics
in the infrared-regulated theory;}

\item {remains finite when the regulation
is taken away. }

\end{itemize}
We will review in a little while how dimensional regularization
can be used to control infrared, as well as ultraviolet,
divergences.

In ``practical" terms, any perturbative
cross section depends on
all the mass scales (collectively, $m$) of the theory in addition to
the relevant kinematic
variables (collectively, $Q$)
$C(m/\mu,Q/\mu,\alpha_s(\mu))$.  An infrared safe quantity
is one whose perturbative expansion
gives finite coefficients in the limit of zero fixed masses, $m$, up to
power corrections in $m/Q$,
\be
C(m/\mu,Q/\mu,\alpha_s(\mu)) = \sum_{n\ge 0} c_n(0,Q/\mu)\, \as^n(\mu)
+ {\cal O}\left[\left(\frac{m}{Q}\right)^p\right]\, .
\label{irsafe}
\ee
with $p>0$, typically an integer.  Depending on the energy range
and the quantity in question,
the mass of a heavy  quark, one well above the
QCD scale, may be treated as an $m$ or as a $Q$.

\section{Introduction to IR Analysis to All Orders}

To explore further the theoretical basis of perturbative analysis
we will work toward the proof of infrared safety for certain
   cross sections, including those for jets in $\rm e^+e^-$
annihilation.  Besides providing results that are interesting
in their own right, this will also provide methods for
the analysis of electron-nucleon and nucleon-nucleon
scattering, as well as for processes involving heavy quarks.

To be specific, we will develop the following necessary ingredients,
sketching, wherever space allows,
arguments that are applicable to all orders
in perturbation theory.

\begin{itemize}

\item {} A review of the treatment of UV and IR divergences in
dimensional regularization;

\item {} A method to identify infrared sensitivity in perturbation
theory: ``physical pictures";

   \item {} A method to identify IR finiteness/divergence: ``infrared
power counting";

\item {} An analysis of cancellation in cross sections: ``generalized
unitarity".

\end{itemize}

The purpose of these discussions will be to  explain the basis of, not
to perform, explicit
calculations.

\subsection{UV and IR divergences in dimensional regularization}

\subsubsection{Navigating the $n$-plane}

We will review a low order example of the use of dimensional
regularization
in due course, but even before that we should clarify how this
technique can
be used to treat both the ultraviolet and infrared sectors of QCD.
The list below outlines in short the logical path from the Lagrangian
of QCD to physically observable quantities using dimensional
regularization.  It also helps to explain why
infrared safety is essential in the comparison of perturbative
calculations
to experiment.

\bea
{\cal L}_{\rm QCD} &\rightarrow& G^{(reg)}(p_1,\dots p_n)\, , \quad
{\rm Re}(n)<4\nonumber \\
       &\rightarrow& G^{(ren)}(p_1,\dots p_n)\, , \quad {\rm
Re}(n)<4+\Delta\nonumber \\
       &\rightarrow& S^{(unphys)}(p_1,\dots p_n)\, , \quad 4<{\rm
Re}(n)<4+\Delta\nonumber \\
       &\rightarrow& \tau^{(unphys)}(\{p_i\})\, , \quad 4<{\rm
Re}(n)<4+\Delta\nonumber \\
       &\rightarrow& \tau^{(phys)}(\{p_i\})\, , \quad n=4\, .
\label{Ltotau}
\eea

First, by means of path integrals or the interaction picture, the
Lagrangian ${\cal L}_{\rm QCD}$
generates a set of rules for perturbation theory, out of which we may
construct the Fourier
transforms of time-ordered products of fields, the Green functions,
$G^{(reg)}(p_i)$
as perturbative integrals over loop momenta.  At this stage, we
regularize the theory
by evaluating the loop integrals in a reduced number of dimensions,
$n<4$,
and by keeping all momenta away from particle poles ($p_i^2\ne m^2$).
If we do this (keeping away from points where subsums of momenta
vanish),
the only poles in the $n$ plane are ultraviolet poles.
These off-shell, regularized Green functions are defined only for $n<4$.
In these terms, renormalization is a systematic procedure for removing
all singularities at $n=4$ in regularized Green functions. The resulting
renormalized Green functions, $G^{(ren)}(p_i)$ may then be analytically
continued
to any $n$ whose real part is less than or equal to $4+\Delta$,
for some $\Delta>0$.

For an infrared finite theory, this would be enough, and
we could safely set $n=4$,  and start evaluating S-matrix elements
and other physical quantities.   For QCD, however, our job
is not done yet, because at $n=4$ the Green functions encounter a
host of infrared, including collinear, singularities.  As a result, its
perturbative S-matrix
elements are hopelessly ill-defined.  Our next step is rather to
continue into
the window $4<{\rm Re}(n)<4+\Delta$.  In this range, we are still
to the left of the ultraviolet divergences of the renormalized theory,
and (as we shall see below) to the {\it right} of infrared divergences,
which  appear in dimensional regularization as poles with ${\rm
Re}(n)\le 4$.
In this infrared-regulated range we can actually take the
limits $p_i^2\to m_i^2$ and construct the S-matrix, for all
quarks and even for the massless gluon.

The S-matrix elements computed for ${\rm Re}(n)>4$ are not physical
of course,  because they have been computed in QCD in more
than four dimensions.
They are not even ``close" to the physical quantities of QCD, as
is manifest by their ubiquitous $1/(n-4)$ poles.  By  contrast, however,
if we can systematically construct quantities $\tau(n)$, which are
at the  same time observable and free of poles at $n=4$,
the $\tau$'s can be returned smoothly to the  physical theory.
The criterion that they lack infrared poles may be taken as an
indication that they are dominated by the short-distance
   behavior of the theory.  These are the quantities to
which  asymptotic freedom can be applied to provide infrared
safe predictions that we may compare to experiment.
Let us see how these ideas are realized for the  simplest
of cases, the total cross section for $\rm e^+e^-$ annihilation
to hadrons.

\subsubsection{The total annihilation cross section}

\begin{figure}[h]
\centerline{\epsfxsize=6cm \epsffile{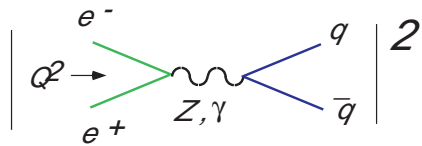}}
\caption{Lowest order $\rm e^+e^-$ annihilation to quarks.}
\label{loann}
\end{figure}

The process
   $\rm e^+e^-\rightarrow$ hadrons already proceeds
at zeroth order in $\alpha_s$ through
annihilation to quarks,
via the photon and the Z:
$\rm e^+e^-\rightarrow$ quark + antiquark,
as in Fig.\ \ref{loann}.
Specializing for the moment to a  virtual photon,
we can readily derive from Fig.\ \ref{loann} the total cross section
for unpolarized annihilation at center of mass energy $Q$,
\be
\sigma^{(0)}_{\rm total}= N_{\rm colors}{4\pi\alpha_{\rm EW}^2\over
3Q^2}\ \sum_f\; Q_f^2\, ,
\label{loepem}
\ee
which is, in fact, a pretty good approximation to the total cross
section below
the Z pole, if we count in the sum over flavors those quark pairs that
can be
produced at a given energy.  In this way, Eq.\ (\ref{loepem}) is a
nice test of quark masses and charges, $Q_f^2=4/9,\, 1/9$, and $N_{\rm
colors}=3$.
  From our previous discussions, we expect this lowest-order perturbative
prediction to make sense for the same reason that lowest-order
perturbative cross sections make sense in QED.
The cross section $\sigma_{\rm total}$ is infrared safe, and
as a result, corrections will appear as a power series in $\alpha_s(Q)$,
which is relatively small once the total energy is well into the
GeV range.  Let's see how this works at first order in $\alpha_s$.

\begin{figure}[h]
\centerline{\epsfxsize=5cm \epsffile{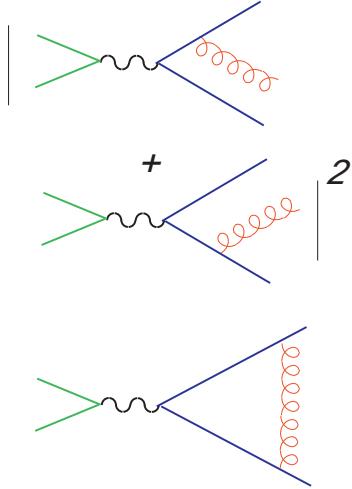}}
\caption{Diagrams for the order $\as$ correction to the annihilation
cross section.}
\label{nloann}
\end{figure}

Diagrams relevant the order-$\as$ corrections to the zeroth-order
cross section are shown in Fig.\ \ref{nloann}.
The first
QCD correction (next-to-leading order, or NLO)  is
\be
\sigma^{(1)}_{\rm total}
=\sigma^{(1,q\bar{q}g)}_{\rm total}
+\sigma^{(1,q\bar{q})}_{\rm total}\, ,
\label{2plus3states}
\ee
   the sum of  contributions
from both three-particle (quark-antiquark-gluon) and two-particle
(quark-antiquark) final states.

Consider first the three-particle final state.
Its contribution to the cross section is
proportional to the lowest order
cross section (\ref{loepem}),
\be
\sigma^{(1,q\bar{q}g)}_{\rm total}=
\sigma^{(0)}_{\rm total}\ \left({\as\over\pi}\right)\,
\left({C_F\over\pi}\right)\, I_3(Q)\, ,
\ee
where $C_F=(N_c^2-1)/2N_c=4/3$ for QCD, and  $I_3(Q)$ is
proportional to the integral over three-particle phase space of the
squared  amplitude
for gluon emission (here in four dimensions),
\begin{eqnarray}
I_3(Q) &=&  (2\pi)\ \int_0^{Q/2} dk\, k\ \int_0^\pi d\theta\
\sin\theta\,
    \left[ {(Q-2k)\over (Q-k(1-u))^2}\right]
\nonumber\\
&\ & \hspace{30mm} \times\;
\left[{[Q - (1-u)k]^2\over k^2(1-u^2)} + {Q(1+u)\over (Q-2k)(1-u)}
\right]\, .
\label{ps3}
\end{eqnarray}
Here $k$ is the gluon
energy, $\theta=\theta_{p_1k}$ is the angle between
the quark and gluon momenta, and $u\equiv \cos\theta$.
In this form, collinear
divergences are manifest at $u\rightarrow \pm 1$,
and infrared (soft gluon) divergences at $k\rightarrow 0$.  Notice that
in this case there
is no collinear divergence at $k=Q$, which corresponds to the quark and
antiquark
becoming parallel, or at zero momentum for the quark or antiquark.

To exhibit the leading divergences, we simplify (\ref{ps3}) to
an approximation relevant to small $k$ and $\theta$:
\begin{eqnarray}
I_{3,\rm IR} =  (2\pi)\,\int_0 {dk\over k}\ \int_0 {d\theta \over
\theta}
\, .
\label{ps3ir}
\end{eqnarray}
We now briefly review how dimensional regularization takes
care of the singularities of these integrals.

\subsubsection{Infrared dimensional regularization in a nutshell}

We can develop the essentials of dimensional regularization
as applied to infrared divergences by
   the following steps,
\begin{eqnarray}
I_{3,{\rm IR}} &=&  (2\pi)\,\int_0 {dk\over k}\ \int_0 {d\theta \over
\theta}
\nonumber\\
&\sim& \int_0 {dk \over k^2} \int_0 {d\theta \over \theta^2}\, \left[
(k\sin\theta)\int_0^{2\phi} d\phi\right]
   \nonumber\\
&=& \int_0 {dk \over k^2} \int_0 {d\theta \over \theta^2}\, \left[
(k\sin\theta)^m\Omega_m\right]_{m=1}
\label{ps3ir2}
\end{eqnarray}
where in the second line we reexpressed the factor $2\pi$ as the
integral over the azimuthal angle.
To motivate the third expression we simply observe that  the azimuthal
angular integral is
just a special case $\Omega_1$ of
$\Omega_m$, the angular integration in polar coordinates for $m+1$
dimensions,
and that the volume element in $m+2$ dimensions is precisely
$(k\sin\theta)^m\Omega_m$, for $m$ any integer
   \footnote{This is easy to verify
inductively, using polar coordinates in  $N$ dimensions to define
cylindrical coordinates in $N+1$ dimensions as an intermediate step.}

With this in mind, we reinterpret the parameter $m$ in Eq.\
(\ref{ps3ir2}),
which in the unregularized theory has the  value unity, as a {\it
complex parameter},
and {\it redefine} $\Omega_m$ as a  complex function $\Omega(m)$
of $m$, restricted
to be the angular volume $\Omega_m$ in $m+1$ dimensions for any integer
value of $m$.
This turns out to be enough to define $\Omega(m)$ uniquely  for all
complex $m$,
\begin{eqnarray}
\Omega_{d} &=& {2\pi^{(d+1)/2}\over \Gamma\left({1\over 2}(d+1)\right)}
\nonumber\\
&=& 2^d\pi^{d/2}{\Gamma\left({1\over 2}d\right)\over \Gamma(d)}\, ,
\label{omegavalue}
\end{eqnarray}
with $\Gamma(z)$ the Euler Gamma function, with
the well-known properties,
\bea
   \Gamma(1/2)&=&\sqrt{\pi},\nonumber\\
z\Gamma(z)&=&\Gamma(z+1),\nonumber\\
   \Gamma(N)&=&(N-1)!\, ,
\eea
   where $N$ is an integer.  With these properties we easily
check the angular volumes for low
dimensions: $m+1=2\ (2\pi)$, 3 $(4\pi)$, 4 $(2\pi^2)$.

   For ${\rm Re}(m)>1$, $I_{3,{\rm IR}}$ becomes finite,
and the soft and collinear divergences appear as
poles at $m=1$, as anticipated,
\begin{eqnarray}
I_{3,{\rm IR}} \sim
\Omega_m\, \int_0{dk\over k}\, k^{m-1}\ \int_0 {{d\theta\over \theta}}\,
\theta^{m-1}
\sim {1\over (1-m)^2}\, .
\label{mpoles}
\end{eqnarray}
This proceedure
   defines a new theory:  QCD in $4+(m-1)$ dimensions,
where $m=1$ is real QCD.

A similar continuation in dimension can always be
found for any observable, since whenever we continue expressions
to $n>4$, we will be able to integrate
over the angular phase space for dimensions above $n=4$.
This, in turn, will regulate the  remaining integrals.
Analogous considerations apply to the ultraviolet divergences
regulated with $n<4$ as in Eq.\ (\ref{Ltotau}) above \cite{dimreg}.

Our notation below will generally be to take $n$ as
the number of dimensions, and to express variations away from $n=4$
through $\varepsilon=2-n/2$.
Again, we emphasize that it is only infrared safe quantities
that can be evaluated at $\varepsilon=0$, where
they return to the  ``real" theory.

For the case at hand, in $n=4-2\varepsilon$-dimensional QCD,
we found double and single poles in $I_{3,{\rm IR}}$.
The full expression is also straightforward to evaluate,
up to terms that vanish for $\varepsilon\to 0$,
and $\sigma^{(1,q\bar{q}g)}_{\rm total}$  is given in this
approximation by
\be
\sigma^{(1,q\bar{q}g)}_{\rm total}(Q,\ve)
=
\sigma^{(0)}_{\rm total}\;
C_F\left({\as\over \pi}\right)\, \left[ {1\over \ve^2} + {3\over 2\ve}
-{\pi^2\over 2} + {19\over 4}\right]\, .
\label{3total}
\ee
Clearly, the three-particle contribution to the total cross section
is highly singular when returned to four dimensions.  Although
unphysical, it is positive, corresponding to the absolute square
of the amplitude for the emission of  a single gluon.

Dimensional regularization, however, also
enables us to compute the two-particle contribution, given by one-loop
vertex and self-energy diagrams.  This calculation is perhaps more
familiar, since it can be carried out in terms
$n$-dimensional loop integrals.  Unlike the three-particle
cross section, it is not an absolute value squared, and need not be
positive.
The result is proportional to the lowest order
cross section,  and is given by
\be
\sigma^{(1,q\bar{q})}_{\rm total}(Q,\ve)
=
- \sigma^{(0)}_{\rm total}\;
C_F\left({\as\over \pi}\right)\, \left[ {1\over \ve^2} + {3\over 2\ve}
-{\pi^2\over 2} + 4\right]\, .
\label{2total}
\ee
Here we  see the same poles as  in the three-particle case, but now
with negative rather than positive signs.  As expected the sum
of the two- and three-particle NLO cross sections
is finite, and the total NLO cross section is very simple,
\be
\sigma^{(1)}_{\rm total}(Q,\ve)
=
\sigma^{(0)}_{\rm total}\;
C_F\left({\as\over \pi}\right)\, \left[  {3\over 4}\right]\, ,
\label{32total}
\ee
a satisfying result that is nonsingular $\ve$, and
can hence be returned smoothly to four dimensions.

Our goal is now to identify classes of cross
sections, involving jets and related observables,
for which we know ahead of time that calculations
like these will provide predictions at  $\ve=0$, that is,
for real QCD.  To do so, we  need first to develop
a better understanding of where poles in $\ve$ come
from,  not only at NLO, but to all orders in the coupling.

\subsection{Physical pictures}

Divergences in perturbation theory can come about in only
two ways: either from the infinite volume of momentum space,
or from infinities in the integrands of individual diagrams.
In renormalized perturbation theory, the former
are eliminated by the systematic construction of counterterms.
The latter, of course, reflect the vanishing of the causal
propagators: $k^2-m^2+i\epsilon=0$, where $m$ may or
may not be zero.  We have kept the infinitesimal
imaginary part of the propagator to remind ourselves that
perturbation theory is defined by integrals in the
complex plane of every loop momentum component.
The integrand of any diagram is thus a meromorphic function
of the $4L$ complex variables of integration in four dimensions, where
$L$ is the number of independent loop momenta.
This will be a very useful observation, as we shall soon see.

To begin with a specific example, consider
the one-loop quark electromagnetic form factor, the sum
of the lowest-order vertex and self-energy corrections to the
electromagnetic current,
\bea
\Gamma_\mu (q^2, \ve) = -  e \mu^\ve \; \bar{u} \;
(p_1) \gamma_\mu v (p_2)\, \rho (q^2, \ve)\, .
\label{emff}
\eea
The function $\r(q^2,\ve)$ is the  correction that appears
times the lowest order annihilation cross section, Eq.\ (\ref{2total})
at NLO for the two-particle final state,
\bea
\rho (q^2,\ve) &=& - {\a_s \over 2\p} C_F \left ( {4 \p \m^2 \over -
q^2 - i \ep} \right )^\varepsilon \;
{\g^2(1-\ve)\g(1+\ve)\over \g(1-2\ve)}\,
   \left\{ {1\over (-\varepsilon)^2} - {3\over 2(-\varepsilon)}   + 4
\right\}\, .
\label{gammaoneloop}
\eea
   Because the electromagnetic current
   is conserved, the ultraviolet divergences in the
   sum of diagrams cancel, and no counterterms are necessary
   in this computation.

The infrared poles of dimensional regularization, which diverge at
$n=4$, require, as pointed out above, that some lines go on-shell lines
for some
real values of the loop momenta.
But on-shell lines at real momenta are not enough.  The momentum
integrations, interpreted as contours in complex
planes must  also be pinched between coalescing poles,
at least one pole on either side of the contour.  If not,
Cauchy's theorem can be used to deform the contour(s)
away from the poles so that the integral can carried out
in such a way that the integrand is always finite.
The requirement that the contours be pinched makes
it possible to identify potential sources
of infrared singularities systematically.

Consider a particular subspace of the momentum space
of a given diagram where
some set of propagators diverge, that is, where some set of lines
are on-shell.   We'll refer to such a subspace as a ``singular surface"
of loop momentum space.
Let us contract all the lines that are not on-shell at the singular
surface
to a point.    The resulting diagram is called a reduced diagram.
The singular surface can give rise to a divergence only if all
the loop momenta of the reduced diagram are pinched between
coalescing singularities at the surface.  We therefore refer
to such a surface as a ``pinch surface".  The existence of a
pinch surface is a necessary condition for the production of an infrared
pole.

It is natural to think of the reduced diagram, all of whose  lines
are on shell as the picture of a process, but in fact
this is not automatic.  In a physical process,
particles travel between points, represented by
the vertices where they begin and end.
In these terms, we can use
a criterion identified by Coleman and Norton \cite{colnor,ste78}
and distinguish
pinch surfaces solely on the basis of their reduced diagrams:
the lines of the reduced diagram must describe a
scattering process in which classical point particles
move between vertices, such that all their motions are mutually
consistent.

To be specific,
for a reduced diagram with lines of momenta $p_k$
to correspond to a pinch surface,  we must be able to assign to
   each of its vertices $i$ a position $x_i$ in space-time
such that if vertices $i$ and $j$ are connected
by line $k$, with $x_i^0>x_j^0$, then
\be
\left( x_i-x_j \right)^\m = (x_i^0-x_j^0)\, v_k^\m\, ,
\label{phys}
\ee
where $v_k=p_k/p_k^0$ is the four-velocity
associated with the on-shell momentum $p_k$.
This corresponds to free propagation between vertices in space-time.
The condition (\ref{phys}) is invariant under
a scale transformations for all points,
keeping momenta fixed:  $x_i\to \lambda x_i$, for all $i$.
For arbitrary $\lambda>1$, the process simply ``gets larger",
and we can think of this as a potential
source of the infrared divergences.

The proof of this assertion is easy, and may be demonstrated adequately
by a simple example, the triangle diagram in Fig.\ \ref{nloann}.
Because the divergences are associated with
momentum dependence in denominators, we simplify further
by suppressing all numerator factors, and just study
the scalar triangle diagram with light-like external momenta, $p_1$ and
$p_2$,
\be
I_\Delta(p_1,p_2)
=
\int {d^n k\over (2\pi)^n}\; {1 \over (k^2+i\ep)\, ((p_1-k)^2+i\ep)\,
((p_2+k)^2+i\ep ) }\, ,
\ee
which, introducing Feynman parameters, can be  expressed in terms of
a single denominator, $D$, that is quadratic in the momenta,
\bea
I_\Delta(p_1,p_2) &=& 2 \int {d^n k \over (2\p)^n} \; \int^1_0 \; {d
\alpha_1 d \alpha_2 d \alpha_3\, \delta
(1-\sum^3_{i=1} \alpha_i) \over D^3}
\nonumber\\
D &=& \alpha_1 k^2 + \alpha_2 (p_1- k)^2 + \alpha_3 (p_2 + k)^2 + i
\ep\, .
\eea
Because $D$ is quadratic, its zeros, and therefore the points
at which the  integrand diverges, are simply the solutions to
a quadratic equation in each momentum component.  A
necessary condition  that each momentum component
be pinched between singularities is
that the two solutions for each $k^\mu$ must coincide, $\m=0\dots 3$,
so that
\bea
{\partial  \over \partial k^\m}\; D(\a_i, k^\m, p_a ) = 0\, .
\label{zeroderiv}
\eea
For the case at hand this leads to the following conditions,
\bea
\alpha_1 k - \alpha_2 (p_1- k) &+& \alpha_3 (p_2 + k) = 0
\nonumber\\
&\Updownarrow&\nonumber\\
\Delta t_1 v_k - \Delta t_2 v_{p_1- k} &+& \Delta t_3 v_{p_2 + k}=0\, ,
\label{deltaxtriangle}
\eea
in terms of times given by $\Delta t_i=\alpha_i\, p_i^0$
   and velocities $p_i/p_i^0$, as noted above.  The second condition
is the statement that if we start at any one of the three vertices,
the total four-distance travelled by going around the loop
is zero.
This is equivalent to the statement that the points are
connected by classical propagation for each of the three
particles.  Now in fact, for specific solutions, one or
two of the times may vanish.  Indeed, whenever a line is
off-shell, its corresponding $\alpha$ parameter must vanish,
or we may deform the $\alpha$ contour and avoid
the point $D=0$ without worrying at all about the momenta.
The points $\a_i=0$, however, are boundaries of the $\a$
integrals.  These boundaries
define the integrals, and therefore cannot be avoided by contour
deformation.
The vanishing of an $\alpha_i$ normally corresponds to
the vanishing of the time of propagation for an off-shell
particle, one contracted to a point in the corresponding
reduced diagram.

There are three solutions for Eq.\ (\ref{zeroderiv}), which illustrate
the generic physical interpretations of $\rm e^+e^-$ annihilation
processes.

\begin{itemize}

\item Collinear to $p_1$: in which two particles of
momentum $p_1-k$ and  $k$ are on-shell and exactly parallel
to one of the external lines.  The third line, with momentum
$(p_2+k)$ is off-shell, and its time of propagation is correspondingly
zero
\bea
k &=&  \zeta p_1\, , \quad \alpha_3 = 0\, , \quad
                \alpha_1 \zeta = \alpha_2 (1-\zeta) \, .
\eea

\item Collinear to $p_2$:
\bea
k &=& - \zeta^\prime p_2\, , \quad  \alpha_2 = 0\, , \quad
                \a_1 \zeta^\prime = \alpha_3 (1-\zeta^\prime)\, .
\eea

\item Soft: for which the solution is $\alpha_1=1$,
while the velocity of the zero-momentum particle is undefined.
There are similar pinch surfaces when the remaining two momenta
vanish.  The physical picture in this case may be thought of
as an infinite-wavelength soft particle coupling to finite momentum
particles at arbitrary points in space-time,
\be
   k^\mu = 0\, , \quad (\alpha_2/\alpha_1) = (\alpha_3 / \alpha_1) = 0\, 
.
\ee

\end{itemize}

\begin{figure}[h]
\hbox{\hskip 2 in {\epsfxsize=6cm  \epsffile{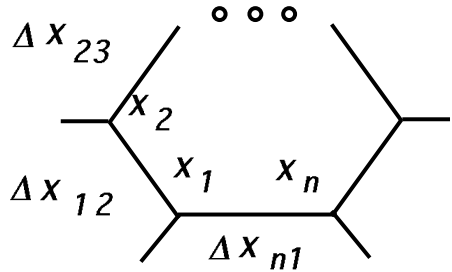}}}
\caption{Loop illustrating the physical condition (\ref{Deltax}).}
\label{loopfig}
\end{figure}

These configurations show that the sources of infrared sensitivity
that we identified above for the high-energy and/or zero
mass limits, collinear and soft, correspond directly to pinch surfaces,
at least for the example that we have examined.
The same analysis, however, can be carried out in just the same way for
an
arbitrary diagram with  arbitrary numbers of loops.
This leads to the generalization of Eq.\ (\ref{zeroderiv}) known as
the Landau equations \cite{Laneqs},
\bea
{\rm for\, all}\, i:\quad  \ell_i^2 = m_i^2,\ {\rm and/or}\
\alpha_i&=&0,
\nonumber \\
{\rm and}\ \sum_{i\; {\rm in}\; {\rm loop} \; s} \; \alpha_i \ell_i
\epsilon_{i s} &=& 0\, ,
\eea
where $\epsilon_{is}$ =1 (-1) for loop momentum $i$ flowing along
(opposite to) momentum
$\ell_i$, and 0 otherwise.
Generalizing Eq.\ (\ref{deltaxtriangle}),
the consistency of the  physical picture associated  with an arbitrary
pinch surface
is ensured by the identity
\bea
\Delta x_{12} + \Delta x_{23} + \ldots + \Delta x_{n1} = 0
\label{Deltax}
\eea
around each loop of the diagram, of the sort in Fig.\ \ref{loopfig}.

The interpretation of pinch surfaces, and hence of the possible
sources of infrared poles, in terms of physical pictures greatly
simplifies
the process of classifying their origin in many cases, particularly
for processes involving hard scatterings.
As we have suggested above, however, satisfying the Landau
equations is still only a necessary, and not sufficient condition
for an infrared divergence.  Clearly, the strength of the singularity
on the pinch surface plays a role as well.  This leads us to
power counting, which provides us with a method for
distinguishing singularities that can give rise to divergences from
those that cannot.

\subsection{Power counting}

As with our analysis of physical pictures, our goal in analyzing the
strength of singularities by power counting is to find necessary
conditions for divergences.   Until we do a specific integral in detail,
it is always possible that an unexpected symmetry
or cancellation between diagrams might eliminate an apparent
singularity.  Indeed, in gauge theories conspiracies between
diagrams are more common that not, because only in gauge-invariant
sums do unphysical modes decouple from physical quantities.
Without going into detail right here, it is worth pointing out that
unphysical modes often give rise to unphysical infrared poles
in individual diagrams \cite{gslab}.

\subsubsection{Example: the triangle with scalar quarks}

We once again turn to the  triangle diagram of  Fig.\ \ref{nloann}
to illustrate the method.  This time, however, we will analyze the
diagram with massless scalar ``quarks" coupled to a massless
``gluon" or photon, taken for simplicity in Feynman gauge,
\bea
\Delta_{\rm IR} = \int_{\rm IR} d^nk
{(2p_1-k)^\mu(-2p_2-k)_\mu \over
\left(-2(p_1^+-k^+)k^--k_T^2+i\ep\right)\,
\left(2(p_2^-+k^-)k^+-k_T^2+i\ep\right)\, (2k^+k^--k_T^2+i\ep)}\, .
\label{scalarqvertex}
\eea
Here we are using light-cone momentum  components,
\bea
k^\pm &=& {1\over \sqrt{2}}\, \left (k^0 \pm  k^3\right)
\nonumber\\
\vec k_\perp &=& (k_1,k_2)\, ,
\eea
with external momenta $p_1^\mu=\delta_{\mu +}$, $p_2^\mu=\delta_{\mu
-}$.

We anticipate on the basis of pinch surface analysis that $\Delta_{\rm
IR}$
is singular when the gluon's momentum $k$ is soft, that is, vanishing
in all four components,
and also when it is collinear to either $p_1$ or to $p_2$.  It is also
clear
that these regions overlap.  We would like to develop a technique that
enables us to verify the behavior of the diagram in these limits,
without
double counting.  Correspondingly, we would like
a means of putting bounds on the integrals
in the presence of the singular points.
We begin by having a look at the limit of soft momentum.

   We can zero in on soft momenta by the following
   rescaling, where we'll be interested in what happens when $\la$
vanishes,
   \be
    \kappa^\mu=\lambda\, k^\mu\, .
\label{softrescale}
\ee
We implement the rescaling by inserting unity in Eq.\
(\ref{scalarqvertex}), in the form
\bea
1_{\rm soft} \equiv \int_0^{\lambda_{\rm max}^2} d\lambda^2 \,
\delta\left(\lambda^2-\sum_\mu k_\mu^2\right)\, .
\label{softunity}
\eea
The argument of the delta function is by no means unique; at present it
is
preferable to choose it as a polynomial
in momentum components and therefore an analytic function of
of the components.
Inserting (\ref{softunity}) into (\ref{scalarqvertex}), we then find the
(noncovariant) expression
\bea
\Delta_{\rm soft} &= - & \int_0^{\lambda_{\rm max}^2} d\lambda^2
\lambda^n\, \int\ d^n\kappa\;
(4p_1^+p_2^- +{\cal O}(\lambda) )\ \lambda^{-2}\,
\delta\left( 1-  \kappa^+{}^2-\kappa^-{}^2-\kappa_\perp^2\right)
\nonumber\\
&\ & \hspace{10mm} \times
{1\over \lambda (-2p_1^+\kappa^-+{\cal O}(\lambda)+i\ep)\,
\lambda (2p_2^-\kappa^++{\cal O}(\lambda)+i\ep)\,
   \lambda^2(2\kappa^+\kappa^--\kappa_T^2+i\ep)}
\nonumber\\
&\sim& - 2 \int_0^{\lambda_{\rm max}} {d\lambda \over \lambda^{5-n}}\
\int d^n\kappa\; {\delta\left( 1-
\kappa^+{}^2-\kappa^-{}^2-\kappa_\perp^2\right)
\over
(-\kappa^++i\ep)\,
(\kappa^-+i\ep)
(2\kappa^+\kappa^--\kappa_T^2+i\ep)}\, .
\label{lambdascaled}
\eea
We have dropped terms that
are higher order in $\lambda$ in both numerator and denominator.
The  integral  is then homogeneous in $\la$, and the
$\la$ integral can be performed to give
a $1/\ve$ pole in $n$ dimensions.   The remaining integral is free of
soft divergences since $\kappa^2$
is fixed at unity.  The soft divergence, where all components
of momentum vanish together, has therefore been isolated
by the scaling (\ref{softrescale}).  This does not
mean that the remaining integral is finite in four dimensions, however,
and we easily verify additional singularities associated
with the presence of collinear divergences
at finite $k^\m$, and arbitrarily close
to the point $k^\mu=0$.  These can be treated in much the
same fashion as for the original diagram, by using the
delta function to do one of the  $\kappa$ integrals,
and then analyzing the analytic structure of the remaining
integral.

The additional collinear divergences appear in the
$\kappa$ integrals as pinches at
the surfaces defined by
\bea
\kappa^\pm &=& 1\, , \nonumber\\
\kappa^\mp &=& \kappa_T^2 = 0\, ,
\label{kappcollinear}
\eea
where we note that the variables $\kappa$ are dimensionless.
These are the soft tails of the collinear regions, and
following a similar analysis we find that they produce
double poles.  For example, for the pinch surface
where $\kappa^+=1$ we may introduce a second scaling,
\be
\kappa^-{}'=\lambda'\, \kappa^-,\ \kappa_T'{}^2 = \lambda'\,
\kappa_T^2, \quad
\lambda'{}^2 = \kappa^-{}^2+(\kappa_T^2)^2\, ,
\ee
and verify that the $\lambda'$ integral, which now controls
the approach to the pinch surface, produces a seond pole in $\ve$.
The remaining integrals are then manifestly finite.
We discover in this way the double poles for the
scalar quark vertex; going through identical steps
in QCD will give the same result.  We now turn to a more general
analysis.

\subsubsection{General pinch surface analysis}

As in the discussion of the  Landau equations, the steps from
the specific example to a general analysis are clear.
Suppose we consider an arbitrary graph $G(Q)$ with
specified external momenta, denoted collectively by $Q$.  Using the
physical picture
analysis, we can in principle identify every pinch
surface $\gamma$ of $G(Q)$.  At any point on
$\gamma$, we may divide the coordinates
of the  loop momentum space into ``internal"
coordinates, $\ell_b$ that specify directions
within the surface $\gamma$, and ``normal" coordinates, $\kappa_a$
that specify directions that move us off the
surface.  For example, for the soft divergence
of the triangle diagram the pinch surface is a point, so that
for this surface all corrdinates $k^\mu$ are normal
coordinates.  For the collinear-$p_1$ pinch surface,
with $p_1$ in the plus direction, the plus component
of the $k$ loop momentum is internal, while the
minus and transverse are normal.

Isolating  the integral of $G(Q)$ near $\gamma$ the same way we treated
the triangle diagram near the soft pinch surface (point), we
write in dimensional regularization
\bea
\hspace{-20mm}
G_\gamma(Q) = \int_\gamma \prod_b d\ell_b
    \int_\gamma \prod_{a=1}^{{\cal D}_\gamma(\ve)} d\kappa_a\
    {\prod_i n_i(\kappa_a,\ell_b,Q) \over \prod_j d_j(\kappa_a,\ell_b,Q)
}\, ,
\label{Ggamma}
    \eea
    where the numerator factors $n_i$ and denominator factors $d_j$
    are polynomials in the normal and intrinsic coordinates, along with
the
    external momenta.  Any Jacobean factors are absorbed
    into the definitions of the volume elements, and the parameter ${\cal
D}_\gamma(\ve)$
    represents the number (possibly noninteger) of normal variables.
Once again,  we insert unity in the  form of a scaling for the  normal
coordinates
(only),
$\kappa_a = \lambda_\gamma\, \kappa_a'$,
\bea
1_\gamma &=& \int_0^{(\lambda_\gamma^{\rm max})^2}\,
d\lambda_\gamma^2\; \delta\left(\lambda_\gamma^2 -
\sum_a\kappa_a^2\right)
\nonumber\\
&=& \int_0^{(\lambda_\gamma^{\rm max})^2}\,
{d\lambda_\gamma^2 \over \lambda_\gamma^2}\;
\delta\left(1 - \sum_a\kappa_a'{}^2\right)
\, .
\label{normalscale}
\eea
As above, the choice of normal coordinates will depend on the pinch
surface.

Under the scaling, both numerators and denominators
have dominant terms in the  $\lambda\to 0$ limit, the
terms of lowest overall powers $N_i$ and $D_j$ of  normal coordinates,
respectively,
\bea
n_i(\kappa_a,\ell_b,Q) &=& \lambda_\gamma^{N_i}\, \left[ \bar
n_i(\kappa_a',\ell_b,Q) +
   {\cal O}(\lambda)\right]
\nonumber\\
d_j(\kappa_a,\ell_b,Q) &=& \lambda_\gamma^{D_j}\, \left[ \bar
d_j(\kappa_a',\ell_b,Q) +
{\cal O}(\lambda)\right]\, .
\label{scalednd}
\eea
In terms of the exponents $N_i$ and $D_j$, we find in this approximation
an overall scaling behavior as we approach the  singular surface, given
by
\bea
G_\gamma(Q) &=& 2 \int_0^{\lambda_\gamma^{\rm max}} {d\la \over
\la^{p_\gamma}}\
\Delta_\gamma(Q)
\nonumber\\
p_\gamma &=& \sum_j D_j + 1 - {\cal D}_\gamma(\ve) - \sum_i N_i\, .
\eea
$\Delta_\gamma(Q)$ is the remaining integral, in which every numerator
and
denominator factor has been replaced by its term of lowest homogeneity
in the $\kappa$'s,
\bea
\Delta_\gamma(Q) =
\int_{{\cal O}(Q)} \prod_b d\ell_b
    \int_\gamma \prod_{a=1}^{{\cal D}_\gamma(\ve)} d\kappa'_a\
    {\prod_i \bar n_i(\kappa'_a,\ell_b,Q) \over \prod_j \bar
d_j(\kappa'_a,\ell_b,Q) }\
    \delta\left(1 - \sum_a\kappa'_a{}^2\right)\, .
    \eea
    On this basis, we estimate the contribution from the pinch surface
    $\gamma$ in dimensional regularization as the  result of  the
    $\lambda$ integral, times whatever the remaining integral
    $\Delta_\gamma(Q)$ gives.

    In general, the integral $\Delta_\gamma(Q)$ will have additional
pinch surfaces,
    of course.  If, however, these pinch surfaces are
    among the original set of surfaces identified for the  graph
    $G(Q)$, we can bound the behavior of $\Delta_\gamma(Q)$
    on this basis, and eventually bound  the entire integral
    by an expression that has some maximal
    number of poles (possibly none) in $\ve=2-n/2$.
    This situation is by no means assured, however, and different
    classes of diagrams require different choices
    of normal coordinates.
    It is by this reasoning that we verify, however,
    the infrared safety of classes of jet and related
    cross sections.  These cross sections are
    in general {\it not} free of pinch surfaces, but
    are defined in such a way that these pinch surfaces
    are not strong enough to produce  poles in $\ve$.

   We note that this analysis
    is simpler for massless quarks.   Organizing poles in
    dimensional regularization is more complicted in the presence
    of quark masses, simply because the masses provide additional
    scales.

\subsubsection{Outline: general analysis for $\rm e^+e^-$ annihilation}

The ingredients above are sufficient to identify the sources of
long-distance behavior in $\rm e^+e^-$ annihilation
with massless quarks quite generally.
   Pinch surface analysis begins with the observation
   that all momentum flows into the hadronic sector
   through a local operator, an electroweak current.
We must identify all physical pictures where particles
emerge from that point, interact and eventually
form a (perturbative) final state.

The reasoning is very simple, keeping in mind that
all particles are taken to be massless, and therefore
propagate at the speed of light.  As a result,  on-shell, finite-energy
particles that emerge moving in {\it different} directions from the
short-distance (``hard")
vertex at which the current acts instantly become space-like
separated and can never interact again to exchange
finite amounts of momentum.
On the other hand particles that emerge from
the current vertex moving in exactly the {\it same} direction may
interact, recombine or split at any time.  The
final state formed in
this way will always consist of sets of exactly collinear particles
for all times.  This is the origin of jets.
In addition, all finite-energy
particles moving in whatever directions
can always be connected by zero-momentum,
infinite wavelength lines.  Thus, although different
jets of particles may not exchange finite
momenta at a pinch surface, they can
exchange color quantum numbers in a nonabelian
theory.

   The reduced diagram for any pinch surface in $\rm e^+e^-$
   annihilation then reduces to:

   \begin{itemize}

   \item{} A set of off-shell-lines
   at the point where the current acts.  We refer to this
   as the hard scattering subdiagram (or function), $H(Q)$.

   \item{} A set of jet subdiagrams, $J_i$ of exactly collinear lines 
that
   emerge from the hard scattering through one or more  finite-energy
   ``jet  lines", which rearrange themselves into any number
   of particles, all moving with exactly parallel momenta.

\item {} A soft subdiagram, $S$, made up of lines whose momenta
   vanish at the pinch surface.  The soft subdiagram may be
   connected to any of the  jet subdiagrams, or to the hard subdiagram
   through zero-momentum lines.

   \end{itemize}

Power counting, carried out as above, may be extended to
pinch surfaces of
semi-inclusive cross sections,  by scaling final-state mass shell
delta functions in the same way as propagators,
which have the same dimensions and homogeneity properties,
$\delta_+(k^2)\ \leftrightarrow\ 1/k^2$.
In this manner, we can show that
in a gauge theory like QCD, cross sections
with fixed  numbers of particles, but with
the same dimension as the total cross section
are at worst logarithmically divergent,
and have up to two poles in $\ve$ per loop \cite{ste78}.
We will refer to such cross sections  as semi-inclusive.
They are simply cross sections for which every
particle in the final state is integrated over
at least part of its phase space.

Space does not allow a detailed proof
of  logarithmic power counting, but we
can think about it in the following way for the
the quark-antiquark production process.
We have already seen that the one-loop vertex
diagram is logarithmically divergent for $n=4$.
Now let's consider what happens when we
attach an extra gluon line to the vertex diagram,
so as to preserve the pinch surface structure.

Adding one additional vector line to such a diagram
increases the number of lines by at most three,
and increases the number of loops by one.
If the line is soft we may choose the normal
variables as all four of the loop momentum
components.  If we attach the soft line at
each end to jet lines,
the denominators of the new jet lines are
linear in the soft momentum, while the denominator of the new soft line
is quadratic.  Scaling the new soft loop along with the
original diagram we find an extra
power $\lambda^{-4}$ from the new denominators
and $\lambda^4$ from the new loop momentum, so
that the power counting remains unchanged.
Adding a jet line leads to the same result,
once the scaling dependence of
momentum factors in the numerator are
taken into account.

\begin{figure}[h]
\hbox{\hskip 2 in {\epsfxsize=9cm  \epsffile{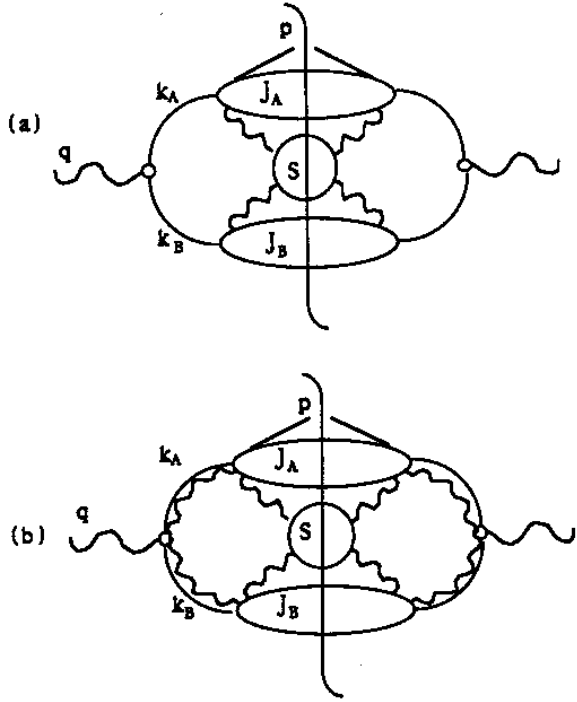}}}
\caption{Representative two-jet reduced diagrams for jet cross section
in cut diagram notation: (a) physical gauge; (b) covariant gauge.
  From Ref.\ \cite{cssrv}}
\label{2jetreduce}
\end{figure}

For a process which like ${\rm e^+e^-}$ annihilation
has a single hard scattering in its reduced diagram,
we may choose normal variables according to
the example shown above.  The reduced diagram
for any such pinch surface will be characterized by
a set of jets, which we may label by index $i$.
Fig.\ \ref{2jetreduce} shows a typical example of a reduced diagram,
with two
jets for simplicity.  The figure  is
presented as a cut diagram, with the perturbative
contribution to the amplitude on the left, and the complex
conjugate amplitude on the right.  The pinch surface
analysis and power counting described above is
readily extended to cut diagrams through the
power-counting correspondence of mass-shell
delta functions and propagators just noted.

The overall power of $\lambda^{\mu(\gamma)-1}$
for  pinch surface $\gamma$ may be thought of
as defining a ``superficial" degree of infrared divergence,
$\mu(\gamma)$
(the extra -1 is compensated by the volume element $d\lambda$) .
The precise graphical and momentum flow configuration of the reduced
diagram
are directly reflected in the value of $\mu(\gamma)$.
At an arbitrary pinch surface, we may identify a number
of relevant integer parameters, associated with the
hard, jet and soft subdiagrams, that provide a
lower bound on the degree of infrared divergence.

\begin{itemize}

\item Some numbers $H_i^{(L,R)}\ge 1$
of lines attach the jets in the amplitude and its
complex conjugate (to the left and right of the final-state
``cut" in a cut diagram).
In the $H$s we count only quark lines and those gluon
lines with transverse polarizations
at the pinch surface.  (In covariant
gauges, ghost lines should also be counted.)
The contributions from
pinch surfaces where one or more jets
connect to a hard scattering by only
unphysical gluon polarizations vanish \cite{gslab}
by the Ward identities of QCD, which ensure
that physical states evolve into physical
states only.

   \item $A_i^{(L,R)}$ unphysically polarized gluons
may, however, appear as additional gluon lines attached to the hard
scatterings.

\item The soft function $S$ may attach to the jet
subdiagram $J_i$, $i=1\dots J$ in an arbitrary fashion, by the
exchange of soft gluons, and in the general case
soft quarks and even ghosts in covariant gauges.
We denote the number of soft gluons, quarks (antiquarks)
and ghosts that connect $S$ with $J_i$ as
$a_i$, $q_i$ and $c_i$, respectively.

\item Some numbers of soft lines ${\cal S}^{(L,R)}$ may
attach $S$ to the left and right hard scattering vertices.
(Such lines are not shown in figure.)

\end{itemize}

The minimal superficial degree of infrared
divergence for an arbitrary $J$-jet pinch surface
in covariant gauges
may be computed in terms of these parameters and is given
in four dimensions by  \cite{ste78}
\be
\mu(\gamma)
= {\cal S}^{(L)}+{\cal S}^{(R)} +
\frac{1}{2}\ \sum_{{\rm jets}\; i=1}^J\
\left \{ H_i^{(L)}+H_i^{(R)}-2+ 2q_i +  c_i + \left[
A_i^{(L)}+A_i^{(R)}+a_i - u^{(3)}_i\right]\right\}\, \, .
\ee
where $u_i^{(3)}$ is the number of three-point vertices in jet $i$
to which soft gluons $a_i$ or jet gluons $A_i$ connect.
In physical gauges, the terms involving the $A_i$s are
absent.

The minimal value of  $\mu(\gamma)$ of a given cut diagram defines the
most
divergent, or ``leading" pinch surfaces (referred to as ``leading
regions" in Refs.\ \cite{ste78,cssrv}).  For negative
$\mu(\gamma)$ we have an overall power divergence;
for $\mu(\gamma)=0$ the overall divergence is
logarithmic, and for positive $\mu(\gamma)$ the pinch
surface does not produce a divergence at all.

For an arbitrary pinch surface relevant to a semi-inclusive cross
section
in $\rm e^+e^-$ annihilation, we easily verify that
the $\mu(\gamma)\ge 0$.  The lower limit, corresponding to
logarithmic divergence, is realized only for reduced diagrams
like the ones of Fig.\ \ref{2jetreduce}.  These are characterized
by quite specific conditions on the variables above.

\begin{itemize}

\item  At most a single quark or physically polarized gluon
         connects each hard scattering vertex to each jet:
$H_i^{(L,R)}=1$.

\item  Arbitrary numbers of scalar-polarized gluons,
with their polarization vectors at the hard
scattering proportional to their momenta, may attach the hard vertices
to the jets in covariant gauges.  These are absent in physical gauges.

\item Soft functions are attached to jets only by soft gluons and at
three-point vertices only.

\item No soft gluons are attached directly to the hard vertex.

\end{itemize}

Our ability to classify pinch surfaces, and to estimate
their degrees of divergence, in this manner is the
basis of effective field theories that systematically separate
the dynamics of hard scattering, jets and soft partons \cite{scet}.

It is worth noting that fermion and/or ghost loops are permitted
internally to the soft function.  In fact, in massless quantum
electrodynamics,
it is through fermion loops that photons interact with each
other, both in the soft subdiagram, and in jet subdiagrams.

Infrared safety requires
that contributions from these well-characterized
   logarithmically divergent regions cancel.
In the next section, we will show how
   this comes about in $\rm e^+e^-$  annihilation.
Our arguments, some of which are new to
these lectures but which are closely related to
those in \cite{ste78,cssrv}, will start with an analysis of
energy flow.

\subsection{Cancellation}

As we have just seen, pinch surfaces for semi-inclusive
cross sections in $\rm e^+e^-$ annihilation are characterized
by fixed numbers of jets.  All interactions within the jet subdiagrams
involve the splitting and/or recombining of collinear lines,
which preserve the total jet four-momenta.
We will now show that if we sum over all final states with the  same
flow of  energy, that is, the same numbers of jets,
these logarithmic divergences will cancel, as long as
the the sum over final states is not weighted by a
function that is itself singular \cite{ste79}.

We begin by introducing an operator that
measures energy flow in an arbitrary state
with specified particles and momenta \cite{kor99},
\bea
{\cal E}(\hat n)\, |k_1\dots k_n\rangle
=
\sum_{j=1}^n\ \omega(\vec k_j)\, \delta^2(\hat{k}_j-\hat{n}) |k_1\dots
k_n\rangle\, .
\label{calEdef}
\eea
To construct the operator $\cal E$, we work in the interaction picture
for the
fields, where creation and annihilation operators are time-independent.
Within this context, the reasoning is quite general for bosons and
fermions,
and we identify the operators by a generic index, $i$.  In these terms,
the energy flow operator is simply
\bea
{\cal E}(\hat n) = \sum_i\,  \int d^3\vec\ell\ a_i^\dagger(\vec \ell)
a_i(\vec \ell)\ \delta^2(\hat \ell -\hat n)\, .
\label{calEadaggera}
\eea
We have chosen to normalize our creation and annihilation operators by
\bea
[ a_i(\vec k),a_j^\dagger(\vec q)]_\pm
=
\delta_{ij}2\omega_k\delta^3(\vec k - \vec q)\, ,
\label{anorm}
\eea
where the $\pm$ subscript indicates commutation for
bosons, anticommutation for fermions.

Using ${\cal E}(\hat n)$, we next construct what we will
refer to as an event shape operator, specified by an arbitrary
function of angle, $f(\hat m,\hat n)$,
\bea
\hat {\cal F}(\hat m) =  Q^{-1}\; \int d^2\hat n\ f(\hat m,\hat n)\
{\cal E}(\hat n) =
   Q^{-1}\; \int d^3\vec\ell\ a_i^\dagger(\vec \ell) a_i(\vec \ell)\,
f(\hat m,\hat \ell)\, ,
\label{calFdef}
\eea
where $\hat m$ represents one or more fixed directions that
we may decide to incorporate into the definition of the weight function
$f$,
and where $Q$ has dimensions of mass.  For $\rm e^+e^-$ annihilation, we
generally take $Q$ to be the center of mass energy.

Any state with definite particle momenta is an eigenstate of $\hat
{\cal F}$,
with an eigenvalue given by the sum of the energies of each particle
times the corresponding weight function,
\bea
\hat {\cal F}(\hat m)\, |k_1\dots k_n\rangle
&=& Q^{-1}\;
\sum_{j=1}^n\ \omega(\vec k_j)\, f(\hat m,\hat k_j)\, |k_1\dots
k_n\rangle
\nonumber\\
&=& f_N\, |k_1\dots k_n\rangle\, .
\label{calFaction}
\eea
The eigenvalue $f_N$ is the total weight of state $|N\rangle\equiv
|k_1\dots k_n\rangle$.

Event shape operators of this kind can be used
to define differential  cross sections in $\rm e^+e^-$ that sum over
all states of a given weight, which we may represent as
\bea
{d\sigma \over df} = L^{\mu\nu}(k_1,k_2)\, {W_{\mu\nu}(q)\over df}
\label{LW}
\eea
for the annihilation of a lepton pair of momenta $k_{1,2}$ into
hadrons.  The cross section is written as the product of
a leptonic tensor, calculable in QED, and a hadronic tensor,
where the indices are associated with the electromagnetic
(more generally electroweak)
currents $j_\mu$ that couple to the quarks.  Using our operator ${\cal
F}$,
we express the hadronic tensor as \cite{kor99}
\bea
\hspace{-15mm}
{dW_{\mu\nu}(q)\over df}
=
\sum_N\ \int d^4x\, \e^{iq\cdot x}\ \langle0| j_\mu(x)|N\rangle\,
\langle N|\delta(f - \hat{\cal F}) j_\nu(0)|0\rangle
\, .
\label{dWdf}
\eea
These are semi-inclusive cross sections
in the sense defined above, and in general the contributions of
states of definite particle number to these cross sections are
logarithmically divergent in the zero-mass limit.

Of course, Eq.\ (\ref{dWdf}) is rather abstract, but we
can represent it more explicitly in terms of
the energy flow operator by using the integral
representation for the  delta function,
\bea
\delta(f - \hat{\cal F})
=
\int_{-\infty}^\infty {d\tau\over 2\pi}
\e^{-i\tau[\, f - Q^{-1} \int d^2\hat n\ f(\hat m,\hat n)\ {\cal
E}(\hat n)\, ]}\, .
\label{taucalE}
\eea
To generate a perturbative expansion, we use
the interaction picture to evaluate the matrix elements in
Eq.\ (\ref{dWdf}),
\bea
&& \hspace{-15mm}
\sum_N\ \langle0| {\rm \bar T}(\e^{-i\int d^4y\, {\cal
L}_I}j_\mu(x))|N\rangle\,
\langle N|\delta(f - \hat{\cal F}){\rm T}(\e^{i\int d^4y'\, {\cal L}_I}
j_\nu(0))\, |0\rangle
\nonumber\\
&& \  =
\int_{-\infty}^\infty {d\tau\over 2\pi}\, \e^{-i\tau\, f}\;
\langle0| {\rm \bar T}(\e^{-i\int d^4y\, {\cal L}_I}j_\mu(x))%
\nonumber\\
\e^{i\tau \int d^2\hat n\ f(\hat m,\hat n)\ {\cal E}(\hat n)} \;
   \; {\rm T}(\e^{i\int d^4y'\, {\cal L}_I} j_\nu(0))\, |0\rangle\, ,
\label{intpict}
\eea
with ${\cal L}_I$ the interaction terms of the Lagrange density.
These matrix elements are evaluated in the free theory, according
to Wick's theorem,
with  T denoting time ordering within the  amplitude, and $\rm \bar{T}$
anti-time ordering in its complex conjugate.

At each order of perturbation theory,
   infrared divergences arise from vanishing phases at large
values of the time variables $y_0$, $y_0'$ for
the interaction Lagrangians.  This claim is adequately
illustrated by a scalar $\phi^3$ coupling, which in terms of creation
and
annihilation operators gives
\bea
\int d^3y\, {\cal L}_I(y_0,\vec y)
&\rightarrow& \int d^3k_1 d^3k_2\; C(k_1,k_2)\,
a^\dagger(k_1)\, a^\dagger(k_2) a(k_1+k_2)
\nonumber\\ && \times
\e^{-iy_0[\omega(\vec k_1)+\omega(\vec k_2) -\omega(\vec k_1+\vec k_2)]}
+ \dots\, ,
\nonumber\\
C(k_1,k_2) &=&
\frac{1}{8(2\pi)^3\omega(\vec{k}_1)\omega(\vec{k}_2)\omega(\vec{k}_1+\ve
c{k}_2)}\, ,
\label{phicubeL}
\eea
where we show a representative term\footnote{With our normalization
\ref{anorm}
for creation and annihilation operators, our scalar field is $\phi(x)=
\int d^3k/[(2\pi)^{3/2}2\omega(\vec{k})]\, [a(k){\rm e}^{-ik\cdot
x}+a^\dagger(k){\rm e}^{-k\cdot x}]$.}.
The time $y_0$ has precisely the interpretation of the time at
which an interaction occurs.  Infrared divergences are produced
only  when interactions from arbitrarily large $y_0$ contribute
(the integrand is just a phase, and is thus finite).
Oscillations in the exponentials of
Eq.\ (\ref{phicubeL}) damp the integrals over $y_0$, {\it unless}
the sum over energies vanishes in the exponent: $\sum \pm\omega_i\to 0$.
It is immediately clear that the combination
   ${-iy_0[\omega(\vec k_1)+\omega(\vec k_2) -\omega(\vec k_1+\vec 
k_2)]}$
vanishes for soft and collinear emissions only.
More generally, the vanishing must occur at a saddle point in
the integrals over momenta, corresponding to pinch surfaces
in the corresponding diagrams \cite{ste95}.

Now, in the absence of the operator involving $\cal E$ in the middle
of Eq.\ (\ref{intpict}),
the large $y_0,y_0'$ dependences of the interaction Lagranians
cancel.  That is, if we make the replacement,
\bea
\e^{i\tau \int d^2\hat n\ f(\hat m,\hat n)\ {\cal E}(\hat n)}
\Rightarrow 1\, ,
\label{toone}
\eea
the exponents cancel for large $y_0, y_0'$.  This is just the
statement that the total cross section is free of IR poles, as
required by unitarity  \cite{kln}.

There are various ways of  showing this result.
One instructive approach uses the  optical theorem,
which relates the total cross section to the imaginary part of the
same matrix elements above with replacement (\ref{toone}),
\bea
\hspace{-20mm}
\sigma_{\rm e^+e^-}^{\rm tot}(Q) &=& {e^2\over Q^2}\ {\rm Im}\, \pi(Q^2)
\nonumber\\
\pi(Q^2) &=& -(1/3) i \int d^4x\ \e^{-q\cdot x}\ \langle0|\, {\rm
T}\left( j^\mu(x)\; j_\mu(0)\right)\, | 0\rangle\, .
\label{optical}
\eea
We now invoke the pinch surface analysis to the
function $\pi(Q^2)$, which is the just
the sum of photon self-energy diagrams.
We recognize that there are no physical processes for
these diagrams at any order of perturbation theory.
Such a process would require a set of jets
to be created at a point and then, without exchanging energy, to
propagate
finite distances in different directions and finally to annihilate.
Such a sequence of events is simply
not possible within a physical picture.

We cannot do quite the same thing for weighted cross sections,
just because the interaction Lagrangian does not commute with
the weight operators $\cal F$ in general.  Nevertheless,
as we have constructed them, the
weight operators $\cal F$ actually do commute with the
interaction Lagrangian at the special momentum configurations
necessary for cancellation
of infrared divergences.
This works because
all $\cal F$  commute with the soft
and ccollinear parts of ${\cal L}_I$.
For example, with the trilinear coupling above,
we can compute
\bea
&&
\left[ \int  d^3 k_1 d^3k_2\; C(k_1,k_2)\, a^\dagger(k_1)\,
a^\dagger(k_2) a(k_1+k_2)\, ,\,
\int d^3\vec\ell\ a_i^\dagger(\vec \ell) a_i(\vec \ell)\,  f(\hat
m,\hat \ell)\; \right]
\nonumber\\
&& \hspace{5mm} =\ \int  d^3 k_1 d^3k_2 a^\dagger(k_1)\, a^\dagger(k_2)
a(k_1+k_2)\nonumber\\
&& \hspace{15mm} \times\;
\left[ -2\omega(k_1+k_2)  f(\hat m,\hat n_{k_1+k_2})
-2\omega(k_1)  f(\hat m,\hat k_1) -2\omega(k_2)  f(\hat m,\hat k_2)
\, \right]\, .
\label{commute}
\eea
The term in brackets on the right vanishes for any collinear set of
momenta,
or when one of the momenta vanishes.
This cancellation of interactions was referred to
as generalized unitarity at the start of this section.  In
the same way, all event shapes functions, or weights,
that are linear in energy times a smooth angular dependence
give matrix elements in which large-time interactions cancel.
Such weight functions are infrared safe.  In addition, once a
differential event shape is shown to be infrared safe, we can take
averages
or moments of it, extremize it with respect to the
variable directions, $\hat m$, and so on.
We now turn to some examples.

\subsection{Jet cross sections and event shapes}
\label{jetshapesec}

The most direct definition of a jet is in terms of
observed energy within a cone, a choice
that we motivated in Sec.\ 2 on
the basis of angular resolution.
Such a cross section can
be implemented in terms of energy flow operators
as above, and is infrared safe.  Actually, some consideration
has to be given to the discontinuous nature of
the corresponding angular
function $f(\hat m,\hat n)$ at the edge of the
jet cone, with $\hat m$ representing
the cone directions and widths.
This does not produce singularities, however,
because divergences are at worst logarithmic,
while the discontinuity is encountered as an endpoint \cite{ste78}.

Any number of variations on the theme of jet cross sections
can be given.  Cluster algorithms are a popular alternative to cone
jets \cite{Ell96}.
In a cluster algorithm, particle  momenta within any final state
are successively combined by searching for the smallest
value of some variable that vanishes when two particles
are collinear.  For example, in the ``$k_T$" or ``Durham" algorithm
applied to $\rm  e^+e^-$ annihilation\footnote{See \cite{jetrunII} for a
$k_T$ algorithm  formulated for hadron-hadron scattering.}, we define
$y_{ij}$ for momenta $p_i$ and $p_j$ by
\bea
y_{ij}= 2Q^{-2}\, {\rm min}\left [(p^0_i)^2,(p^0_j)^2\right] \left
(1-\cos \theta_{ij} \right )\, .
\eea
We start with the pair of particles that has the minimum value
for $y_{ij}$.  These two momenta
are combined into a single momentum according to any continuous rule
that amounts to simple addition when the angle $\theta_{ij}$
vanishes.  The algorithm is then repeated, until all remaining
pairs of momenta satisfy the condition
\be
y_{ij}>y_{\rm cut}\, ,
\ee
for an arbitrary, and adjustable, value of the parameter $y_{\rm cut}$.
The sets of particles whose momenta are grouped in this way
constitute the jets for that particular value of $y_{\rm cut}$.
The smaller $y_{\rm cut}$ the more collimated are the
jets, and of course, the number of jets generally increases
as $y_{\rm cut}$ decreases.   This illustrates that
neither the grouping of particles into jets nor even the number
of jets in a given final state is unique.

Uniqueness is not
necessary, however, for the concept of
the number of jets
to be useful.  What is necessary is that it be possible to
compute the distributions of jet number  self-consistently
in perturbation theory.  An infrared safe
$N$-jet cross section  appears
at lowest order as an $N$-parton production process.
This provides a basic  physical interpretation.
Corrections to this approximation
associated with different, more complex quantum
histories are inevitable, but  calculable.

Labeling a given state as two-jet or three-jet
says almost as much about the definition of a jet as
about the final state.  It is therefore also useful to analyze final
states more directly through weights called event shapes \cite{shaperv}
that can take
on continuous values.  The thrust \cite{far77} is the prototype event
shape.

The thrust quantifies the deviation of a final state from a perfect
configuration of two ``back-to-back" jets, by projecting
momenta on an arbitrary axis $\hat m$, and then
maximizing with respect to $\hat m$:
\be
T= {1\over Q}{\rm max}_{\hat m}\ \sum_i |{\hat m}\cdot {\vec p}_i|\, .
\label{Tdef}
\ee
When $t\equiv 1-T$ approaches zero, the final state consists of two
very narrow jets.
Another interesting relation, quite easy to verify using light cone
variables, is
that for narrow jets, $1-T$ is proportional to the sum of the  squares
of the invariant masses of the  two jets, computed from particle
momenta within the hemispheres defined by the plane normal to
the thrust axis,
\be
t=1-T \sim {M_{J_1}^2+M_{J_2}^2\over Q^2}\, .
\label{tMreln}
\ee
Equivalently, if we define the $z$ direction in the direction of jet
$J_1$
(chosen arbitrarily), and we define a standard rapidity for each
particle momentum
relative to that axis,
\be
\eta_k=\frac{1}{2}\, \ln \left(k^+\over k^-\right)\, ,
\ee
then, neglecting particle masses, the deviation of the thrust from
unity may also be written as
\bea
t = {\sqrt{2}\over Q} \left[ \sum_{i\in H_R}\ k_i^-  + \sum_{i\in H_L}
k_i^+ \right]
= {1\over Q}\sum_{{\rm all}\ i}\ k_{iT}\; \e^{-|\eta_i|}\, .
\eea
Starting from this form, we can generalize the event shape $t=1-T$ to
\bea
e_a
= {1\over Q}\sum_{{\rm all}\ i}\ k_{iT}\; \e^{-|\eta|(1-a)}\, ,
\label{angdef}
\eea
where $e_0$ is $t=1-T$, and $e_1$ is called the ``jet broadening"
\cite{broad}.
As a class, the $e_a$'s have been named ``angularities"
\cite{ber03a,ber04}.

\begin{figure}
\hbox{ \hskip 3 cm {\epsfxsize=12cm \epsffile{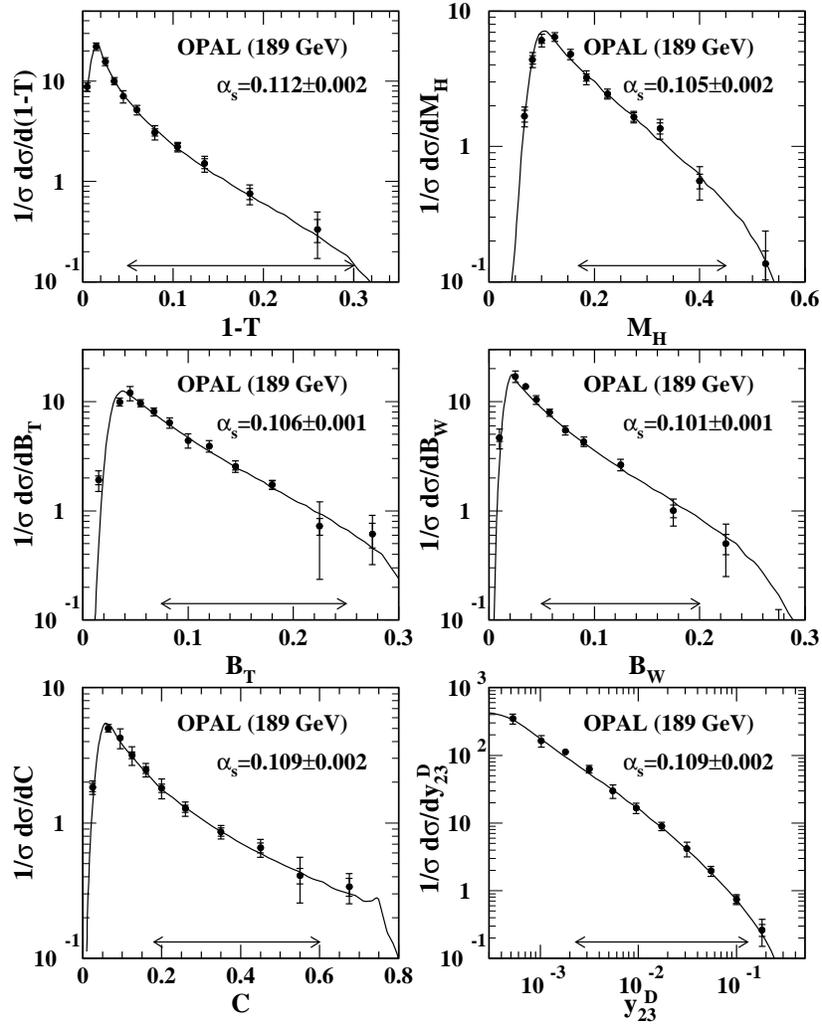}}}
\caption{Distributions of various event shapes
as measured by the OPAL Collaboration.}
\label{opalfig}
\end{figure}

Figure \ref{opalfig} shows a variety of event shapes as measured
at the LEP collider \cite{opal}, including the thrust, and variations
of the broadening.
We  notice  that except for the two-jet limit near $1-T=0$, for example,
the event shapes are relatively slowly varying functions.  The arrows
in each panel
show the range over which the OPAL Collaboration decided to compare
fixed
orders in perturbation theory to experiment to determine the value
of the strong coupling at the Z mass, $\as(M_Z)$, with the results
shown.
Near the two-jet limit, however, the distributions peak, and then
show a rapid decrease.  This behavior cannot be reproduced by
a few orders in perturbation theory, but requires an all-orders
analysis.

   As we've already observed,
   both the thrust distribution $d\sigma/dT$ and
   its Mellin moments $\int_0^{T_{\rm max}} dT T^N d\sigma/dT$ are
infrared safe.
The rapid turnover near $T=1$ may be traced to a singularity in the
thrust distribution there, which appears at ${\cal O}(\as)$ as
\bea
\hspace{-45mm}
{1\over \sigma_0}\;
{d\sigma(Q)\over  dT} &=& \delta(1-T) - 2C_F\ {\as(Q) \over \pi}\ \left
[{\ln(1-T)\over 1-T}\right ]_+
+\cdots \nonumber\\
{1\over \sigma_0}
\int_0 dT T^N\; {d\sigma(Q)\over  dT} &=& 1 - C_F\ {\as(Q) \over \pi}\
\ln^2 N+\cdots\, .
\label{largeN}
\eea
Here the ``plus" distribution
   $[...]_+$ is associated with the sum of virtual-gluon and real-gluon
   corrections.  It will be defined below.

  From a physical point of view, taking
   $T\to 1$ is like taking energy resolution to zero in QED, and we
   expect the cross section to vanish in that limit, simply because
   radiation is required when charged particles are accelerated.  In QED
this is
   generally more a matter of principle than of practical concern,
   because the coupling is small.  The strong coupling even at the
   Z mass is not so small, however, and the corrections of (\ref{largeN})
   are not only significant, they dominate the cross section near the
   peak.  This is characteristic of event shape distributions, for which
   the bulk of events are precisely in region where corrections
   from every order in perturbation theory are large.
This sets the stage  for a discussion of resummation in QCD, which
enables us to bring such corrections under control.  We will
come back this issue later, after a discussion of applications of
perturbative
QCD to deep-inelastic scattering.

\section{Deep-inelastic Scattering and Factorization}

In this section, we review the basic structure of deep-inelastic
scattering, showing how QCD at once justifies and generalizes
the parton model.  We will introduce the method of factorization,
which enables us to use infrared safety, and hence asymptotic
freedom in DIS and other hard scattering processes with
hadrons in the initial state.  In this section, we will deal primarily
with low-order examples.  A brief discussion of what is involved
in proofs of factorization theorems will be given in the final
section.

We'll begin with a quick review of DIS in the parton model,
followed by a statement of its generalization to QCD in
the form of factorized structure functions.  We go on to
sketch the procedure of calculating perturbative corrections
to the structure functions, including the derivation of
one-loop evolution kernels (splitting functions).
All these calculations are elementary, but they serve as a prototype
for the analysis from which we derive predictions for
both new-physics signals and QCD backgrounds and tests.

\subsection{DIS and the parton model}

\begin{figure}
\hbox{\hskip 2 cm {\epsffile{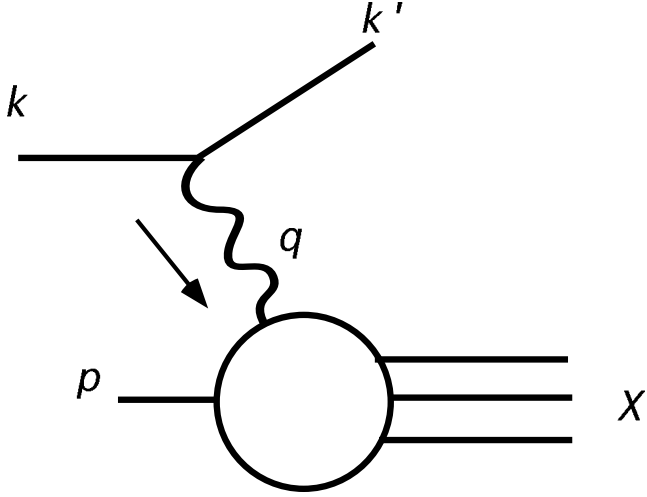}}}
\caption{DIS kinematics.}
\label{disamp}
\end{figure}

\begin{figure}[t]
\centerline{\epsfxsize=5cm \epsffile{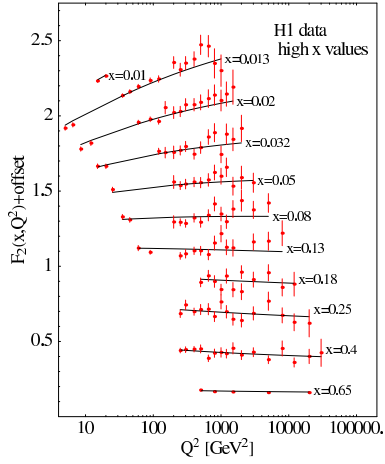}}
\caption{H1 data for the structure function $F_2$.}
\label{h1highxfig}
\end{figure}

As introduced in Sec.\ 1, deep-inelastic scattering (DIS)
is the reaction $e(k) + N (p) \to e (k^\prime) + X_{\rm hadronic}$,
where the momentum transfer $q=k-k'$ is large.
Fig.\ \ref{disamp} represents the basic process.
The corresponding (unpolarized) cross section can be broken down into
the product of leptonic and hadronic tensors, in just the
same way as $\rm e^+e^-$ annihilation,
\bea
d \s = {d^3 k^\prime \over 2s | \vec{k}^\prime |} \; {1\over (q^2)^2} \;
L^{\m\n} (k, q) W^{\gamma N}_{\m\n} (p, q)\, ,
\label{LWdis}
\eea
with the leptonic and hadronic parts given by lowest-order QED and
by hadronic matrix elements, respectively,
\bea
L^{\m\n} &\equiv& {e^2 \over 8 \p^2} tr \left[ \rlap{/}{k} \gamma^\m
\rlap{/}{k}^{\; \prime}
\gamma^\n \right] \nonumber\\
W^{\gamma N}_{\m\n} &\equiv& {1\over 8 \p} \sum_{{\rm spins}\ \s} \;
\sum_X  \langle N (p, \s) \mid J_\m (0) \mid X \rangle\;
\nonumber \\
&\ & \hspace{20mm} \times \langle X \mid J_\n (0) \mid N (p, \s )
\rangle
   (2 \p )^4 \delta^4 (p_X - q - p )\, .
\label{LthenWdis}
\eea
The superscript $\gamma N$ reminds us that  we are neglecting
Z boson exchange.
The symmetries of the strong interactions, already mentioned in
connection with Eq.\ (\ref{sigmaasFs}) for the DIS cross section, lead
to
the following decomposition of $W^{\mu\nu}$  in terms of invariant
tensors,
\bea
W_{\m\n}^{\gamma N} &=& - \left ( g_{\m\n} - {q_\m q_\n \over q^2}
\right )
W_1^{\gamma N}(x, q^2)
\nonumber\\
&& \hspace{20mm}+ \left (p_\m + q_\m \left ( {1\over 2x} \right )
\right )
\left (p_\n + q_\n\left  ({1\over 2x} \right ) \right ) W_2^{\gamma N}
(x, q^2 )\, ,
\label{Wtensors}
\eea
which are related to the dimensionless structure functions of Eq.\
(\ref{sigmaasFs}) by
\bea
F_1^{\gamma N}(x,Q^2)\equiv W^{\gamma N}_1,
\quad\quad F_2^{\gamma N}(x,Q^2)=p\cdot q\; W_2^{\gamma N}\, ,
\label{WtoF}
\eea
where again, the ``scaling" variable is given by
$x = - {q^2 / 2 p \cdot q} \equiv {Q^2 / 2 p \cdot q}$,
and as noted above, scaling is the statement that
the structure functions depend only on $x$, not on $Q$.
Contemporary data in Fig.\ \ref{h1highxfig} from the  H1 experiment at
DESY shows that for a wide range of $x$ and $Q$
scaling is an experimental, albeit approximate,
statement of fact.
Recalling the discussion of Sec.\ 2, the parton model
cross section for electron-nucleon DIS
at a given $x$ and $Q^2$ is given by the product of the
lowest-order electromagnetic ``Born" cross section
for electron-quark scattering, times the  probability
distribution for quarks (of flavors $f$)
within the hadrons at momentum fraction $x$,
\bea
   d \s^{(\ell N )} (p, q) = \sum_f \int^1_0 d \x \; d \s_{\rm
Born}{}^{(\ell f)} (\x p,q) \f_{f/N } (\x)\, .
   \label{pmdis}
   \eea
In the parton model, the quark distributions, $\f_{f/N}(x)$
are independent of $Q$.

We can break down the parton model cross section (\ref{pmdis})
into leptonic and hadronic tensors, with the latter given by
\bea
W_{\m\n}^{\gamma N} &=& {1\over 8 \p}
   \int {d^3 p' \over (2\p)^3 2 \o_{p^\prime}}
Q_f{}^2\, {\rm Tr}\, [\gamma_\m {\rlap{\,/}p}^{\; \prime} \gamma_\n
\rlap{\,/}p ]
\ (2\p)^4 \delta^4  (p^\prime - \x p - q )\, .
\label{pmW}
\eea
  From here we identify
structure functions in the parton model,
following the general analysis (\ref{LthenWdis}), (\ref{Wtensors})
and (\ref{WtoF}), which gives
\bea
F_2^{\gamma N} (x) &=& \sum_f \int^1_0 d \x\, F_2^{\gamma f(0)} (x/\x)
\f_{f/N} (\x )
\nonumber\\
F_1^{\gamma N} (x) &=& \sum_f \int^1_0 \; {d \x \over \x}\,
F_1^{\gamma f(0)} (x / \x)
\f_{f /N} (\x )\, ,
\label{pmF}
\eea
where the Born structure functions are found
by the replacing the external hadron by external quarks.
In Eq.\ (\ref{pmF}),  the Born structure functions are then
\bea
2 F_1^{\gamma f(0)} (z) = F_2^{\gamma f(0)} (z) = Q_f{}^2 \delta
(1-z)\, .
\label{Fqborns}
\eea
Inserting (\ref{Fqborns}) into (\ref{pmF}), we find
\bea
2x F_1^{\gamma N} (x) = F_2^{\gamma f(0)} (x) = \sum_f Q_f\, x\,
\phi_{f/N}(x)\, ,
\label{cgreln}
\eea
which is known as the Callan-Gross relation \cite{calgro}.
The structure functions are thus proportional to the
quark distributions in the parton model, and hence to each other.
As anticipated in Sec.\ 2, this relation depends on the quark spin,
and is thus analogous in significiance to the $1+\cos^2\theta$
dependence of jets
in $\rm e^+e^-$ annihilation, providing direct insight into parton
spin.

   \subsection{Factorization in DIS}

We now discuss how QCD incorporates the successes of the
parton model.   The challenge is to use asymptotic
freedom in quantities like structure functions.
Structure functions clearly
depend upon hadronic structure and
are therefore not infrared safe, so we have no
hope of computing them in perturbation theory.
Factorization is the term that describes a
general approach to this problem.  It is applicable in one form or
another
for any cross section that includes a short-distance
subprocess.

The particular form of factorization
that we will discuss is adapted to generalize the
parton model,  and is
sometimes referred to as ``collinear" factorization.
We'll mention some other possibilities later,
but collinear factorization is the best
understood and most highly developed.
For the remainder of this section, we'll use the terms factorization
and collinear factorization interchangeably.

A factorized cross section will be a product or
convolution of two pieces: an infrared
safe part (short-distance), which is calculable in perturbative,
QCD, and an infrared sensitive part
(long distance), which although not perturbatively calculable,
is measureable and universal among a class of factorizable
processes.   For DIS and related
cross sections, the infrared factors are parton distributions, or PDFs.
The same general method of factorization can also be applied to
elastic scattering amplitudes at large momentum transfer
\cite{qcdform}.

\begin{figure}[h]
\centerline{\epsfxsize=10cm \epsffile{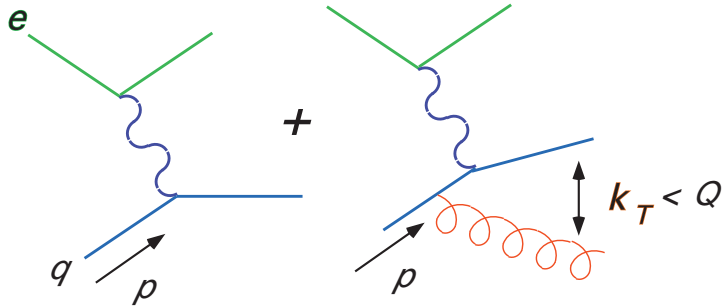}}
\caption{Gluon emission at lowest order in a DIS process.}
\label{nlokt}
\end{figure}

The form of a  factorized cross section is very close to
the parton model, but unlike the parton model, the
short-distance functions are not normalized uniquely
   by partonic cross sections.
Beyond the lowest order in perturbation theory, the
split between short- and long-distance is not unique.
In fact, the parton model would be exact if there were
a lower limit to the lifetime of virtual states in QCD,
corresponding to an upper limit in the energy deficits
of such states.
If this were the case, once the momentum transfer $Q$
were large enough, quantum interference between QCD
and electroweak scattering would disappear.
This is not the case, of course, as illustrated by Fig.\ \ref{nlokt},
which shows the emission of a single gluon just before
a quark absorbs a photon.  The lifetime of the quark/gluon
virtual decreases with an increase in the transverse momentum of the
gluon, $k_T$, which is limited only by the available energy.
On the other hand, for very small $k_T$, the diagram
develops collinear singularities.  It is these two
regimes we need to factorize.
These issues will be clarified by
studying the first nontrivial order.

\subsection{Field theory corrections}

To see how factorized cross sections work, we study
DIS not of a hadron, but of a quark.  What does this
mean in perturbation theory?  To compute a quark
structure function, we must
work in the (dimensionally) regulated theory, with
the number of dimensions $n>4$.   Although, as
emphasized above, this is not the real theory of QCD,
infrared safe quantities computed in this theory
can be returned to four dimensions, and
incorporated into realistic QCD. So let's go ahead.

\subsubsection{Factorization for quark DIS}

Let us begin with the statement of factorization, in this
case for the structure function of a quark, given by the convolution of
a short-distance, infrared safe function, $C$, which generalizes
the parton model, and an infrared-sensitive function $\phi$,
\begin{eqnarray}
F_2^{\gamma q}(x,Q^2) &=&
\sum_{i=q\bar{q},g}
\int_x^1 d\xi \; C_2^{\gamma i}
\left( {x\over \xi},{Q\over \mu},{\mu_F\over \mu},\alpha_s(\mu)\right)
\ \phi_{i/q}(\xi,\mu_F,\alpha_s(\mu))
\nonumber\\
&\ &\hspace{15mm}
\nonumber\\
&\equiv&  C_2^{\gamma i}
\left( {x\over \xi},{Q\over \mu},{\mu_F\over \mu},\alpha_s(\mu)\right)
\otimes \phi_{i/q}(\xi,\mu_F,\alpha_s(\mu))\, .
\label{quarkF2fact}
\end{eqnarray}
The first equality defines the convolution indicated by the second.
The sum is over all flavors of quark and antiquark, as well as the
gluon.
Radiation may produce any of these partons in the
initial state, and the transition from long to short distances may occur
for any parton.  We  note, of course, that when $i=g$, the coefficient
function
must be at least order $\as$, since  the gluon must fluctuate back
to a  quark pair to absorb a photon.
The variable $\mu_F$ is the factorization scale, which will separate
short- and long-distance dynamics, while $\mu$ is the renormalization
scale.  For our purposes, it is convenient to identify these two
scales, $\mu_F= \mu$, although we need not do so.
We also often encounter the direct choice  $\mu=Q$, which emphasizes
the perturbative calculability of $C$,
\be
F_2^{\gamma q}(x,Q^2) = {C_2^{\gamma i}
\left( {x\over \xi},\alpha_s(Q)\right)}\ \otimes
\ {\phi_{i/q}(\xi,Q^2)}\, .
\label{F2qfact}
\ee
Eq.\ (\ref{quarkF2fact}), however, illustrates the general statement of
factorization.

\begin{figure}[h]
\centerline{\hskip 12mm \epsfxsize=16cm \epsffile{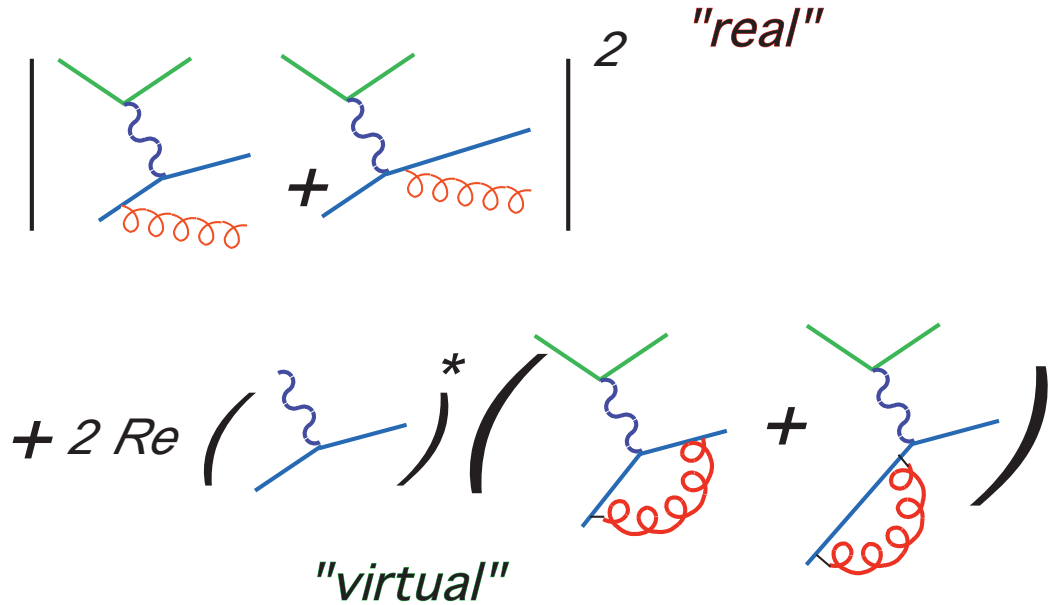}}
\caption{NLO diagrams for $F_2$.}
\label{disoneloop}
\end{figure}

The zeroth order expansion of the structure function $F_2$ is found
from the
Born graph of elastic quark scattering, while
NLO contributions to Eq.\ (\ref{quarkF2fact}) are
calculated from the diagrams of Fig.\ \ref{disoneloop} in $n$
dimensions.
We suppress,
but hopefully remember,  dependence on the
variable $\ve=2-n/2$, which will appear as poles in
the infrared-sensitive factors of this expression.

At lowest order (no QCD), the distribution of a parton in
a parton is a delta function,
\bea
\phi_{q/q'}^{(0)}(\xi)=\delta_{qq'}\; \delta(1-\xi)\, ,
\eea
which expresses the obvious fact that in the
absence of any interactions, the  quark's momentum fraction
is  conserved.  Then, comparing the factorized
structure functions of Eq.\ (\ref{F2qfact}) to the Born quark
structure functions in Eq.\ (\ref{Fqborns}), we confirm that
at zeroth order the factorized coefficient functions
are the Born structure functions (\ref{Fqborns})
\bea
C^{\gamma q(0)}_{1,2}\left({x\over \xi}\right)
   = Q_q^2\ \delta(1-x/\xi)\, ,
\eea
so that at lowest order the factorized quark structure function reduces
to
\begin{eqnarray}
{F_2^{\gamma q\, (0)}(x,Q^2)} &=& \sum_i\; \int_x^1 d\xi \;{C_2^{\gamma
i\, (0)}
\left( {x\over \xi}\right)}
\ {\phi^{(0)}_{i/q}(\xi)}
\nonumber\\
&\ &
\nonumber\\
&=&  {Q_f^2}\ \int_x^1 d\xi\;  {\delta(1-x/\xi)}\
{\delta(1-\xi)}
\nonumber\\
&\ &
\nonumber\\
&=&  {Q_f^2\ x\, \delta(1-x)}\, ,
\label{0orderF2q}
\end{eqnarray}
exactly as in the parton model.  We are now ready to incorporate QCD
corrections.

\subsection{Quark DIS to one loop}

Moving on to NLO (often termed one loop,
even though gluon emission at this order is really
tree level), as shown in Fig.\ \ref{disoneloop},
we recognize that we are going to have contributions from two- and
three-particle
phase space.  Recalling our discussion of
infrared and collinear divergences, we anticipate
that there will be at least some cancellation of poles in $\ve$.
The treatment of these technical issues is simplified by
the introduction of plus distributions.  Like delta functions, plus
distributions are generalized functions, and are defined by their
integrals with smooth functions.  The definitions of
the plus distributions we'll need are,
\bea
\int_0^1 dx\ {f(x)\over (1-x)_+} \quad &\equiv& \int_0^1 dx\
{f(x)-f(1)\over (1-x)}
\nonumber\\
{\int_0^1 dx\ f(x){\left(\ln(1-x)\over 1-x\right)_+}}
&\equiv&
{\int_0^1 dx\ \left(\, f(x)-f(1)\, \right)\ {\ln(1-x) \over (1-x)}}\, ,
\label{plusdistdef}
\eea
with an obvious to generalization to higher powers of logs and other
numerator functions.
In factorized cross sections, the role of $f(x)$ will
be played by products of parton distributions and other smooth
functions.  At first order, the $f(x)$ term
corresponds to the emission of a  real gluon, with momentum fraction
$1-x$, while the $f(1)$ term will correspond to virtual gluon
corrections, with elastic kinematics for the quark.
A very special distribution, one we will encounter shortly, is
is the  evolution kernel or splitting function for quark-to-quark
evolution,
\begin{eqnarray}
{P_{qq}^{(1)}(x) = C_F\ \left[{1+x^2\over 1-x}\right]_+}\, .
\label{dglap}
\end{eqnarray}
For us $P_{qq}$ will have the interpretation
of the probability of gluon emission per unit log transverse momentum.

Let us now relate the first two orders (LO, NLO) on the
left- and right-hand sides of Eq.\ (\ref{quarkF2fact}).
The left-hand side is defined uniquely by
the diagrams in (regularized) perturbation theory, but we
need a criterion for splitting
the result from the left into a coefficient function
and parton distribution on the right.  Such a criterion is usually
called a factorization scheme.   In any scheme,
we can expand the right-hand side in orders of $\as$,
\begin{eqnarray}
F_2^{\gamma q_f}(x,Q^2) = C_2^{(0)}\otimes \phi^{(0)}
   + {\alpha_s\over 2\pi}\
C_2^{(1)}\otimes \phi^{(0)}
   + {\alpha_s\over 2\pi}\; C_2^{(0)}\otimes \phi^{(1)} + \dots\, ,
\label{expandright}
\end{eqnarray}
where we introduce the notation $f(\as)=\sum_n(\as/2\pi)^nf^{(n)}$.
Into this template we can insert the explicit calculation of
the left-hand side, given for $F_2$ by
\begin{eqnarray}
{F_2^{\gamma q_f}(x,Q^2)} &=&  Q_f^2\ x\; \Bigg\{\ \delta(1-x)
\nonumber\\
&\ &
\nonumber\\
&\ & \hspace{-10mm}  + \
{\alpha_s\over 2\pi}\
{ C_F\left[ {1+x^2\over 1-x} \left(\ln{(1-x)\over x}\, -\,
\frac{3}{4}\right)
+{1\over 4}\;
(9+5x)\right]_+}
\nonumber\\
&\ &
\nonumber\\
&\ & \hspace{-10mm}   + \
{\alpha_s\over 2\pi}\ {
   C_F\;
\left[{1+x^2\over 1-x}\right]_+}\
\left(4\pi \mu^2{\rm e}^{-\gamma_E}\right)^{\ve}
\int_0^{Q^2} {dk_T^2\over k_T^{2+2\ve}}\ \Bigg\}\ +\cdots\, .
\label{quarkF2nloexpand}
\end{eqnarray}
and for the correction to the Callan-Gross relation by
\begin{eqnarray}
{2xF_1^{\gamma q_f}(x,Q^2)} &=&
   F_2^{\gamma q_f}(x,Q^2) - C_F\, {\alpha_s\over \ \pi}\; x^2\
\, .
\label{FLnlo}
\eea
In Eq.\ (\ref{quarkF2nloexpand}), the dots indicate additional
$\ve$-dependence that vanishes in four dimensions.
The remaining integral over $k_T$ refers to the
transverse momentum of a virtual or real gluon
in the diagrams of Fig.\ \ref{disoneloop}.
This term illustrates the interpretation of the
evolution kernel (\ref{dglap}) as a probability
for gluon emission per unit logarithm of $k_T$,
in four dimensions.  At low, $k_T$, however,
we see that infrared regulation is required,
as confirmed by the necessity of
$\ve<0$, i.e., $n>4$ to define the integral.

The calculation that leads to Eq.\ (\ref{quarkF2nloexpand}) is
straightforward, and
a clear exposition of all details, for both quarks and gluons
may be found in \cite{emw}.   The two- and three--body final states
both contain double poles in $\ve$, which, however, cancel in the
inclusive
hadronic cross section.
The factor in the final line contains the remaining infrared
sensitivity, which can be identified as a collinear divergence.
The factor $\mu$ is here the renormalization scale in
the dimensionally regularized theory and the
factor $4\pi{\rm e}^{-\gamma_E}$,  with $\gamma_E$ the Euler
constant,  is characteristic
of all one-loop integrals.  As expected, despite the cancellation of
double poles, this
structure function cannot be returned to four dimensions.
We are now
ready to discuss schemes for choosing the infrared
safe coefficient functions.

\subsubsection{Factorization Schemes}

As long as $\ve<0$, we are free to perform the
$k_T^2$ integral in Eq.\ (\ref{quarkF2nloexpand}),
\be
\int_0^{Q^2} {dk_T^2\over k_T^{2+2\ve}}
=
{1\over -\ve}\; Q^{-2\ve}\, .
\ee
The pole generated by this integral,
times the DGLAP splitting function, is the only term in Eq.\
(\ref{quarkF2nloexpand})
that must be absorbed into the definition of $\phi^{(1)}$ in Eq.\
(\ref{expandright}).
In the minimal subtraction, or $\overline{\rm MS}$, scheme for the
quark distribution in $n=2-2\ve$ dimensions, this is exactly what we do,
\begin{eqnarray}
{\rm \overline{\rm MS}:} \hspace{20mm}
{\phi_{q/q}^{(1)}(x,\mu^2) =
   \left(4\pi\mu^2{\rm e}^{-\gamma_E}\right)^\ve\; P_{qq}(x)\
\int_0^{\mu^2}
{dk_T^2\over k^{2+2\ve}_T}}\, ,
\label{msbarphi1}
\end{eqnarray}
with $\mu$ the factorization scale.
The advantage of this choice is its technical simplicity and generality.
The coefficient function at one loop is easy to find,
\bea
C_2^{(1)}(x)_{\overline{\rm MS}} =  P_{qq}(x) \ln(Q^2/\mu^2)
+ \mu{\rm -independent}\, ,
\label{Cmsbar}
\eea
where all of its factorization scale dependence
is specified by the splitting function.  Of course, other choices
of scheme are possible, by shifting other finite terms from
the coefficient functions into the distributions.
Closer to the parton model than the $\overline{\rm MS}$ scheme
is the DIS scheme, which is defined to absorb all
corrections into the distribution for deep-inelastic scattering,
\begin{eqnarray}
{\rm DIS:} \hspace{20mm}
x\, Q_q^2\, \phi^{(1)}_{q/q}(x,\mu^2) = F_2^{\gamma q\, (1)}(x,\mu^2)\,
.
\label{disscheme}
\end{eqnarray}
In this case,
\bea
C_2^{(1)}(x)_{{\rm DIS}} = x\, P_{qq}(x) \ln(Q^2/\mu^2)
+ 0\, .
\label{Cdis}
\eea
Notice that even in DIS scheme, the coefficient inherits the
same dependence on the factorization scale as in the $\overline{\rm MS}$
scheme.
Other schemes, closer to other processes are possible, but nowadays,
most calculations are presented in a minimal scheme.

\subsubsection{Using the regulated theory: parton distributions for
real hadrons}

As we have repeatedly observed,
IR-regulated QCD is not the real QCD, but since it differs only
in its infrared sensitivity, we should be able to use infrared safe
functions like the $C_2^{\gamma q}$'s even in four dimensions,
where we cannot calculate parton distributions for real hadrons,
at least perturbatively.  This will enable us to get parton
distributions from real hadrons.
Here is how it works:

\begin{itemize}

\item  Compute $F_2^{\gamma q}$, $F_2^{\gamma G}$ (for weak bosons as
well).

\item  {Define a factorization scheme; find infrared safe $C$'s}.

\item  {Use these coefficients in the full theory for hadron $N$,}
\bea
F_2^{\gamma N}(x,Q^2) &=& \sum_{a=q_f,\bar q_f,G}\; \int_x^1 d\xi \;
C_2^{\gamma a}
\left( {x\over \xi},{Q\over \mu},{\mu_F\over \mu},\alpha_s(\mu)\right)
\ \phi_{a/N}(\xi,\mu_F,\alpha_s(\mu))
\nonumber\\
&\equiv& C_2^{\gamma a} \otimes \phi_{a/N}\, .
\label{disfact}
\eea

\item  Measure the physical structure functions
$F_i$ (not neglecting W and Z boson exchange and neutrino scattering);
   then use the known $C$'s to derive the full set of $\phi_{a/N}$ for
$q=q,\bar{q},G$.

\end{itemize}

Compared to our discussion above,
multiple flavors and structure functions essentially complicate
the technicalities, but
not the logic.   Of course, the final step is a highly nontrivial
exercise in undoing the convolutions that relate distributions
to observables.
Assuming we can carry out this procedure, however,
the result is a collection of parton distributions,
$\phi_{a/N}(\xi,\mu^2)$, for all partons $a$ (quarks and
antiquarks of various flavors, and the gluon) in
the nucleon (proton and neutron).
The parton distributions can now be used in
any process that factorizes into distributions
times perturbative hard-scattering functions.
What we have not explained yet is how to combine
different choices of $\mu_F$.  To that end, we turn
to evolution.

\section{Evolution}

   Evolution enables us to determine the $Q^2$-dependence
of structure functions, and more than that, to extrapolate
parton distributions from one energy scale to another.
It is the cornerstone of predictions at high energy both
for QCD and new physics signals.

   For the purposes of this discussion we set the
factorization scale equal to the renormalization scale, $\mu_F=\mu$,
so we write
$C_a\left(x/\xi,Q^2/\mu^2,\alpha_s(\mu)\right)$.
We now choose $\mu=Q$, so that $F_2$, for example, is given
for hadron $A$ by
\bea
{F_2^{\gamma A}(x,Q^2)} = \sum_a \int_x^1 d\xi \;{C_2^{\gamma a}
\left( {x\over \xi},1,\alpha_s(Q)\right)}
\ {\phi_{a/A}(\xi,Q^2)}\, .
\label{muisQ}
\eea
The larger is $Q$, the better the perturbative expansion,
\bea
C_2^{\gamma a}\left( {x\over \xi},1,\alpha_s(Q)\right)
=\sum_n \left({\alpha_s(Q)\over\pi}\right)^n\,
C_2^{\gamma a}{}^{(n)}\left( {x\over \xi}\right)\, .
\eea
To make this work for us, we only need the
distribution at $\mu=Q$.

Assuming that we have determined our PDFs at
some scale $Q_0$, evolution makes it
possible to determine them at any other scale
$Q$, and hence to determine the
structure functions (for example) at any other $Q$.
Our only requirement is that both $Q$ and $Q_0$
should be large enough that both $\as(Q)$
and $\as(Q_0)$ are small.
In particular, extrapolations to arbitrarily
large energies are possible.

We can illustrate these ideas most easily  by studying a ``nonsinglet"
structure function,
defined as the difference between proton and neutron structure functions
\bea
F_i^{\gamma \rm NS} = F_i^{\gamma p}  - F_i^{\gamma n}\, ,
\eea
which may be factorized in terms of a single nonsinglet parton
distribution,
\bea
{F_2^{\gamma \rm NS}(x,Q^2)} &=& \sum_a \int_x^1 d\xi \;{C_2^{\gamma
\rm NS}
\left( {x\over \xi},{Q\over\mu},\alpha_s(\mu)\right)}
\ {\phi_{\rm NS}(\xi,\mu^2)}\, .
\label{nsstruc}
\eea
In this expression, the contributions of gluons and antiquarks have
cancelled,
and the nonsinglet distribution is a sum over quark flavors.
In fact, these cancellations are relevant only beyond one loop;
at one loop $C_2^{\rm NS}=C_2^{\gamma N}$.

It is now convenient to take the Mellin transform of Eq.\
(\ref{nsstruc}),
defined by
\bea
\bar f(N)=\int_0^1dx\ x^{N-1}\ f(x)\, .
\label{moment}
\eea
This has the advantage of reducing  convolutions of this kind to
products,
\bea
f(x) = \int_x^1 dy\ g\left({x\over y}\right)\ h(y) \rightarrow
\bar f(N)=\bar g(N)\ \bar h(N+1)\, .
\label{convolprod}
\eea
In our case, we find
\bea
{\bar F_2^{\gamma \rm NS}(N,Q^2)}
=
{\bar C_2^{\gamma \rm NS}
\left( N,{Q\over\mu},\alpha_s(\mu)\right)}\  {\bar \phi_{\rm
NS}(N,\mu^2)}\, ,
\label{F2moment}
\eea
where in a slight shift of
notation we define
$\phi_{\rm NS}(N,\mu^2)\equiv\int_0^1 d\xi \xi^{N}\phi(\xi,\mu^2)$ here.

We now observe that
   ${\bar F_2^{\gamma \rm NS}(N,Q^2)}$ is physical, so that, as in Eq.\
(\ref{physzero}),
it cannot depend on the factorization/renormalization scale, $\mu$,
\bea
    \mu{d\over d\mu}\ {\bar F_2^{\gamma \rm NS}(N,Q^2)} = 0\, .
\eea
Applied to Eq.\ (\ref{F2moment}), this means that variations in
the coefficient function and in the parton distribution must compensate
each other,
\bea
\mu{d\over d\mu} \ \ln  \bar \phi_{\rm NS}(N,\mu^2) &=&
-\gamma_{\rm NS}(N,\alpha_s(\mu))
\nonumber\\
\gamma_{\rm NS}(N,\alpha_s(\mu)) &=&  \mu{d\over d\mu} \ln  \bar
C_2^{\gamma \rm
NS}
\left( N,\alpha_s(\mu)\right)\, .
\label{separation}
\eea
The function $\gamma_{\rm NS}(N,\alpha_s(\mu))$
plays the role of a separation constant.  It is a function
of $\as$ and $N$ only, because these are the only variables
that $C$ and $\phi$ hold in common.  In particular,
it has an infrared safe expansion in $\as$.
Thus, to get the $\gamma$'s we need to know only $C$'s, which
we can get from the infrared regulated theory as above.
The anomalous dimensions $\gamma(N)$ found in this
way, however, enable us to determine the $\mu_F$ dependence
of the parton distributions.  For any factorized cross section,
we may choose $\mu_F$ at the hard scale in the problem,
and make predictions at high energy based on observations at lower
energy.

In summary, the $Q$-dependence of factorized cross sections
is determined by perturbation theory through evolution.
This procedure was how we found out that QCD is the ``right"
theory for the strong interactions.  It is also the way QCD predicts
physics at as-yet unseen scales.

\subsection{Evolution at one loop}

We now have all the pieces necessary to derive the consequences
of asymptotic freedom and factorization in deep-inelastic scattering.
According to (\ref{separation}), we can compute the anomalous dimensions
directly from the coefficient functions, noting that the answer is the
same in either (or any) factorization scheme,
\begin{eqnarray}
\gamma_{\rm NS}(N,\alpha_s) &=&
\mu{d\over d\mu}\ {{\ln  \bar C_2^{\gamma \rm NS}
\left( N,\alpha_s(Q)\right)}}
\nonumber\\
&\ & \hspace{-45mm}
= \mu{d\over d\mu}\
\left\{\; {
(\alpha_s/2\pi)\ \bar P_{qq}(N) \ln(Q^2/\mu^2)
+
\mu\ {\rm indep.}}\; \right\}
\nonumber\\
\ \nonumber\\
&\ & \hspace{-15mm}
= -{\alpha_s\over \pi}\ \int_0^1 dx\;x^{N-1}\; P_{qq}(x)
\nonumber\\
\ \nonumber\\
&\ & \hspace{-15mm}
   = -{\alpha_s\over \pi}\, C_F\ \int_0^1 dx\; \left[
\left(x^{N-1}-1\right)\, {1+x^2\over 1-x}
\right]
\nonumber\\
\ \nonumber\\
&\ & \hspace{-15mm}
= {\alpha_s\over 2\pi}\, C_F\ \left[\; 4\sum_{m=2}^N{1\over m} -{2\over
N(N+1)} +1\; \right]
\nonumber\\
&\ & \hspace{-15mm} \equiv
{\alpha_s\over \pi}\ \gamma_{\rm NS}^{(1)}\, .
\label{gammanoneloop}
\end{eqnarray}
We have used the explicit form of the nonsinglet
splitting function, Eq.\ (\ref{dglap}).  These nonsinglet anomalous
dimensions are the same as those for the full structure functions,
to lowest order.
To confirm the integrals the
identity
\be
(1-x^k)/(1-x)=\sum_{i=0}^{k-1}x^i
\ee
may be helpful.

Having done the work, it's just a small step to some extraordinary
physics
predictions \footnote{As the written form of these lectures were being
prepared, the 2004 Nobel Prize honoring the discovery of asymptotic
freedom was awarded.  The field theoretic explanation of scaling and
scale
breaking, in much the same manner as presented here, was the original
link of QCD to experiment and
remains one of its strongest links to this day}.
Going back to Eq.\ (\ref{separation}), the very anomalous dimensions
computed from the NLO coefficient functions control the factorization
scale dependence of the parton distributions,
\bea
\mu{d\over d\mu} \ { \bar \phi_{\rm NS}(N,\mu^2)} = -\gamma_{\rm
NS}(N,\alpha_s(\mu))\
{ \bar \phi_{\rm NS}(N,\mu^2)}\, ,
\label{phievol}
\eea
The solution to this equation is an exponential, where the exponent
is  an integral over the  scale appearing in the running coupling,
\begin{eqnarray}
{\bar \phi_{\rm NS}(N,\mu^2)}
&=& {\bar \phi_{\rm NS}(N,\mu_0^2)}
\ { \exp\left[\;  -{1\over 2}\; \int_{\mu_0^2}^{\mu^2}\;
{d\mu'{}^2\over \mu'{}^2}\ \gamma_{\rm NS}(N,\alpha_s(\mu))\;
\right]}\, .
\label{phisoln}
\eea
Of course, we can evaluate this expression for any theory,
once we determine the anomalous dimensions and the
running of the coupling as above.  If  (the generic case) the
theory is not asymptotically free, the
coupling increases rather than decreases with
$\mu'$, on the basis of its one-loop
beta function.  At the same time the anomalous
dimensions, the analogs of $\gamma^{(1)}_{\rm NS}(N,\as)$,
remain positive.   Once the coupling is large, of course, it may
no longer be possible to use the one-loop anomalous dimensions,
and it is still possible that the coupling constant
might approach a finite constant (a fixed point) due to
higher-order terms in the beta function.  This, however, would
predict violations of scaling by a power in $\mu$, and hence in $Q$
\cite{chm}.
In contemporary terms, for such theories the parton distributions
decrease
rapidly in $\mu$, and
hence the moments of the structure functions decrease very strongly
with $Q$.
In physical terms, a decrease in the moments corresponds
to a rapid shift of the parton distributions from larger to smaller $x$
as $Q$
increases.
This was the paradox of scaling in 1973: all known field
theories seemed to predict too much radiation for partons
to retain their momentum fractions $x$ at large $Q^2$.

In QCD, of course, asymptotic freedom changes this.
The QCD coupling (\ref{alphamumuprime}) decreases with  scale.
We can perform the  $\mu'$ integral in the exponent of (\ref{phisoln})
easily, to get
\bea
{\bar \phi_{\rm NS}(N,Q^2)}\
&=& {\bar \phi_{\rm NS}(N,Q_0^2)}
\ {
\left({\ln(Q^2/\Lambda_{\rm QCD}^2)
\over
\ln(Q_0^2/\Lambda_{\rm QCD}^2)} \right)^{-2\gamma_N^{(1)}/b_0 }}\, ,
\label{losoln}
\end{eqnarray}
which is also often written as
\begin{eqnarray}
{\bar \phi_{\rm NS}(N,Q^2)}\
&=& {\bar \phi_{\rm NS}(N,Q_0^2)\
\left({\alpha_s(Q_0^2)\over
\alpha_s(Q^2)} \right)^{-2\gamma_N^{(1)}/b_0 }}\, .
\label{losolnalpha}
\end{eqnarray}
Scale breaking has not disappeared, but it is now logarithmic,
and hence very mild once $Q$ is significantly larger than
the scale $\Lambda_{\rm QCD}$.   This establishes
both the consistency of QCD with the parton model,
and a constellation of predictions for improvements
to the parton model, which have been confirmed
in the intervening years.

The scaling violation predictions of QCD are easiest to
compute via moments of the parton distributions as above.
For numerical  evaluation, however, it is often more convenient
to invert the moments, which takes the
simple rate equation for the parton density in (\ref{separation}) back
to a convolution
\be
\mu{d\over d\mu} \ {  \phi_{\rm NS}(x,\mu^2)} = \int_x^1 {d\xi \over
\xi}\
P_{\rm NS}(x/\xi,\alpha_s(\mu))\ { \bar \phi_{\rm NS}(\xi,\mu^2)}\, ,
\label{nsconvol}
\ee
in terms of the very same splitting function,
\be
\int_0^1 dx\; x^{N-1}\ P_{qq}(x,\alpha_s) = -
\gamma_{qq}(N,\alpha_s)+\dots
=\int_0^1 dx\; x^{N-1}\ P_{\rm NS}(x,\alpha_s) + \dots\, ,
\ee
where again, the nonsinglet and quark-quark splitting functions
are the same to lowest order.
Equation (\ref{nsconvol}), commonly referred to as the DGLAP evolution
equation \cite{dglap}, also provides a nice physical interpretation of
evolution,
as a sequence of radiation processes.  At each such step,
the parton  momentum fraction $\xi$ is degraded to $x<\xi$,
in an essentially probabalistic manner.

\subsection{Beyond nonsinglets}

The generalization of  these results to the full  system of
parton distributions is simple, at least in principle.  As long as
the sequential radiation processes that drive evolution
involve only gluons, the evolutions of individual quark
flavors are independent.   The (singlet) structure functions for
individual hadrons, however, receive contributions from
fermionic as well as gluonic radiation, which couples
the evolutions of all the quarks, antiquarks and the gluon.
Compared to Eq.\ (\ref{nsconvol}), the only change is that
the nonsinglet splitting function is replaced by a matrix $P_{ba}$,
each element of which describes the production of
partons of flavor $b$ from those of flavor $a$,
\bea
\mu{d\over d\mu} \ {  \phi_{b/A}(x,\mu^2)} &=&
- \sum_{b=q,\bar q,G}\int_x^1 {d\xi \over \xi}\
P_{ba}(\xi,\alpha_s(\mu))
\ { \phi_{a/A}(\xi,\mu^2)}\, ,
\label{sconvol}
\eea
or in moment space
\bea
\mu{d\over d\mu} \ { \bar \phi_{b/A}(N,\mu^2)} &=&
-\sum_{b=q,\bar q,G}\
\gamma_{ba}(N,\alpha_s(\mu))
\ { \bar \phi_{a/A}(N,\mu^2)}\, ,
\label{smoment}
\eea
in terms of a matrix of anomalous dimensions, $\gamma_{ba}(N)$.
Of special interest is the low-$x$ behavior of $P_{gg}(z) \sim 1/z$,
which corresponds to $\gamma_{gg}(N)\sim 1/(N-1)$.  For
$N\sim 1$ moments, then, the effects of evolution can
be large, simply because large anomalous dimensions
raise the logarithms corresponding to those in (\ref{losolnalpha})
above to a high power.  The strong scale-breaking effects of such terms
in the anomalous dimensions are
clearly seen in the data of Fig.\ \ref{H1lo}.
The systematic study of low-$x$ evolution is
a subfield in its own right, which starts with the
famous BFKL equation \cite{bfkl}, which organizes the
leading logarithms in $x$ at each order of $\as$.  An introduction can
be
found in Ref.\ \cite{Ell96}, and a summary of recent highlights
in Ref.\ \cite{dis03}.

\subsection{Summary}

Here is a summary our discussion of factorization and evolution.
In the parton model, $\phi_{a/A}(x)$
denotes the density of partons $a$ with momentum fraction $x$,
a distribution that is assumed to be quantum-mechanically
independent of the hard scattering
at momentum transfer $Q$, and hence may be  treated as an independent
probability.  In QCD, $\phi_{a/A}(x,\mu)$ represents the
same density, but only of partons with transverse momentum $\le \mu$,
as illustrated by the integral of Eq.\ (\ref{quarkF2nloexpand}).
It is only these partons whose production may be considered
incoherent with the hard scattering.

   If there {\it were} a maximum transverse momentum
$Q_0$ for partons in the nucleon, $\phi(x,Q_0)$
would freeze for $\mu\ge Q_0$, and the theory would
revert to the parton model above that scale.
This is never the case, however, in a renormalizable
field theory, and scale breaking measures the change in the density as
the maximum
transverse momentum increases.
   Of course, the structure functions and
cross sections that we compute still depend on our choice of
$\mu$ through uncomputed higher orders in $C$ and  evolution.
At present, evolution for DIS is routinely implemented
fully up to NLO, which requires
order $\as^2$ in the splitting functions $P_{ab}$.
This formalism can successfully describe the behavior
of structure functions over a wide range of  $x$,
as illustrated by Figs.\ \ref{h1highxfig} and \ref{H1lo}.
\begin{figure}[h]
\centerline{\epsfxsize=7cm \epsffile{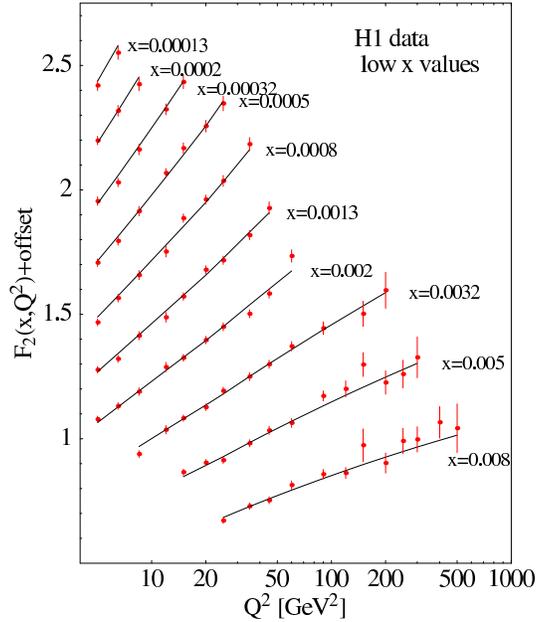}}
\caption{DIS data at low $x$, compared to
fits from \cite{cteq6}.}
\label{H1lo}
\end{figure}
The full splitting functions up to $\as^3$ have just become
available within the  past year \cite{as3split}, and the development of
a phenomenology at NNLO is underway \cite{nnlophen}.

\section{Applications and Extensions}

In this final section, we will touch on a (very
incomplete) subset of the
generalizations of the jet and DIS analyses
we have just described, including factorization for hadron-hadron
scattering, the basis of all-orders factorization proofs
and of resummations, and a few current areas of progress
relevant to the LHC program.

\subsection{Factorized hadron hadron cross sections}

Space does not allow an extensive description of
the application of factorization and evolution to hadron
hadron scattering \cite{cssrv,Ell96}, but even a cursory description
shows the utility of these techniques.
The factorization formalism can be
extended to the semi-inclusive production of heavy particles
or systems of jets, labeled $F$, of mass $Q$.
Indeed, the decay of a system of heavy particles, be it a
top quark pair or SUSY signal, usually results in hadronic
jets.  An essential element of new-particle searches is to
tease the signals of these decays out of the QCD and  electroweak
background.

The production of final-state
$F$ requires the short-distance collisions of a parton
from each hadron, and the same pinch surface and power
counting analysis as for $\rm e^+e^-$ annihilation and DIS
shows that the entire leading power in $Q$ comes
from only a single (physical) parton from each hadron \cite{cssrv}.
The essential requirement for factorization
is once again that we sum over states involving
the emission of soft particles and/or the rearrangement
of collinear radiation.
It is important that the definition
of the semi-inclusive final state not involve any scales
much smaller than $Q$, the mass of the observed final
state, because the largest corrections will generally
be an inverse power of the smallest such scale.

A factorized cross section for
hadron-hadron scattering describes the production
of final states by a direct generalization of DIS
factorization, as a convolution of two parton distributions
with a  hard-scattering function,
\begin{eqnarray}
\sigma_{AB\rightarrow F(Q)}(Q) &=&
\sum_{a,b=q,\bar q,G}\ \int_0^1 dx\, dy\,
\phi_{a/A}(x,\mu)\ \phi_{b/B}(y,\mu)
\,  \hat \sigma_{ab\rightarrow
F(Q)}(xp_A,yp_B,Q,\mu,\alpha_s(\mu))
\nonumber\\
&\ &
\hspace{35mm} \times
\theta(xy p_A\cdot p_B - Q^2)\, ,
\label{hhfact}
\end{eqnarray}
where we have exhibited a theta function to emphasize the requirement
that
the partonic system has sufficient energy to produce system $F$
of mass $Q$.
As usual, the hard-scattering function
$\hat \sigma$ begins at lowest order with a
Born cross section, and depends on the choice of
$\mu$ through uncomputed higher orders.

An illustrative
   ${\cal O}(\alpha_s)$ (NLO) hard-scattering cross section is the
   quark-antiquark annihilation cross section
$\hat \sigma_{\bar q q}$ for the Drell-Yan process
   (${\rm q\bar{q}} \rightarrow \gamma\  {\rm or\ Z}
\rightarrow \ell^+\ell^-$), given in $\rm\overline{MS}$ by
\begin{eqnarray}
\hspace{-15mm}
\hat \sigma_{\bar q q}^{(1)}
&=& \sigma_{\rm Born}(Q^2)\ C_F\;  \left({\alpha_s(\mu)\over
\pi}\right)\ \Bigg\{
P_{qq}(z)\, \ln\left({Q^2\over\mu^2}\right) + 2(1+z^2)\;
\left[{\ln(1-z)\over 1-z}\right]_+
\nonumber\\
&\ & \quad -{\left[(1+z^2)\ln\, z\;\right]\over (1-z)} +
\left({\pi^2\over 3}-4\right)\;
\delta(1-z)\Bigg\}\, ,
\label{dynlo}
\end{eqnarray}
where $z=Q^2/x_ax_bS$ with $\sqrt{S}$ the total center of mass
energy squared, and where
   $\sigma_{\rm Born}$ describes the electroweak annihilation of $q\,
\bar q$.
We see again a plus distribution, whose presence
reflects the cancellation of real and virtual soft gluons.

A  perturbative prediction for this process, based on factorization, is
compared to data at high energy ($\gamma$ and Z),
is shown in Fig.\ \ref{dycdffig} from CDF at the Tevatron,
The peak due to the Z is clearly visible, and indeed this was
the way the Z was first directly seen.
Also shown is data for the forward-backward
asymmetry associated with the parity-violating
predictions of the standard model, and some
extenstions of it.  Notice that this asymmetry, with
or without a new physics signal, is
calculable precisely because it is part of the hard
scattering function.
\begin{figure}[t]
\centerline{\epsfxsize=11cm \epsffile{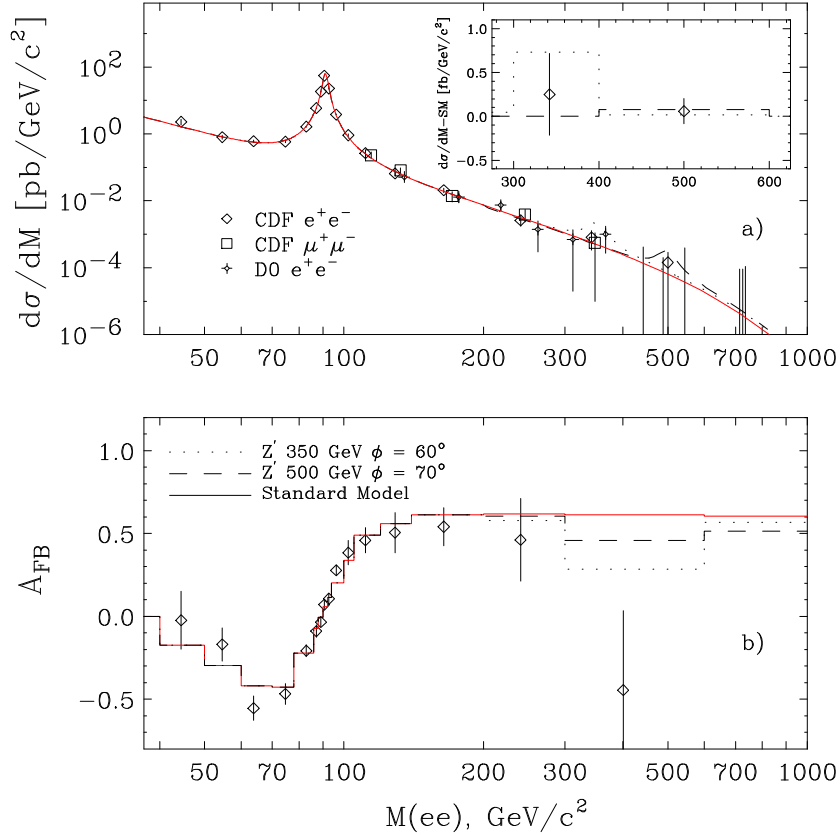}}
\caption{Drell-Yan data from the Tevatron \cite{dycdf}.}
\label{dycdffig}
\end{figure}
Such tests, for signals of new states or violations of symmetries,
are the prototypes for the detection of new physics in hadronic
scattering.

The successes and remaining uncertainties of predictions based on
factorized perturbation theory are well illustrated by the data for
inclusive high-$p_T$ jet cross sections, based on formulas that are
analogous
to those for Drell-Yan, but
(much, much) more elaborate even at NLO \cite{nlojets}.  The data track
perturbative predictions over many orders of magnitude, but
even between the two Fermilab experiments, differences in
analysis of the still-preliminary data \cite{run2data} show that there
is still
work to be done.
\begin{figure}[h]
    \centerline{\hbox{ \hspace{0.2cm}
      \includegraphics[width=6.7cm,height=6.5cm]{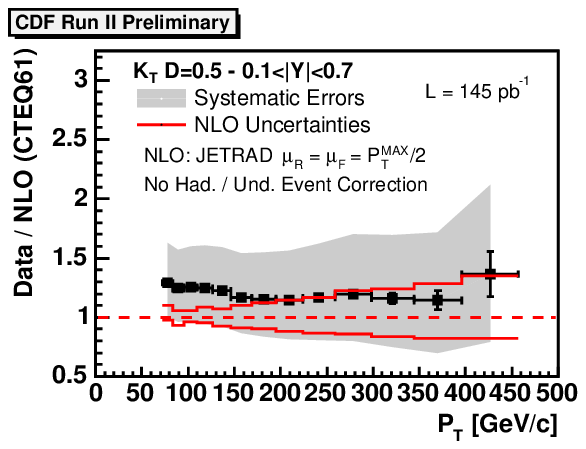}
      \hspace{0.3cm}
      \includegraphics[width=6.7cm,height=6.5cm]{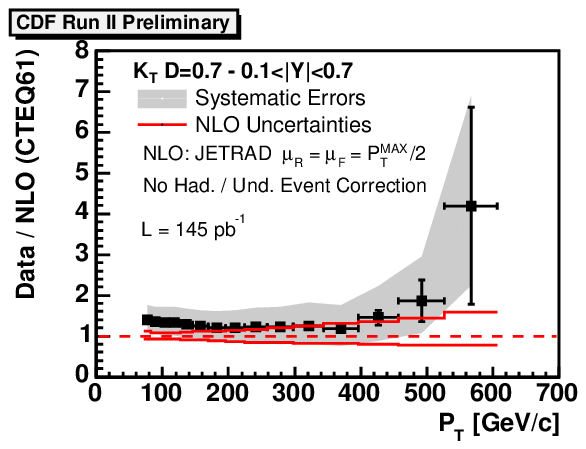}
      }
    }
\caption{Recent preliminary CDF jet data \cite{run2data}.  The data is
presented here as the ratio of the difference between data and the
NLO prediction, normalized to the prediction.  As the following figure
shows, the absolute value of the data is changing by many orders of
magnitude over
this range.}
\end{figure}

\begin{figure}[h]
    \centerline{\hbox{ \hspace{0.2cm}
      \includegraphics[width=6.7cm,height=6.5cm]{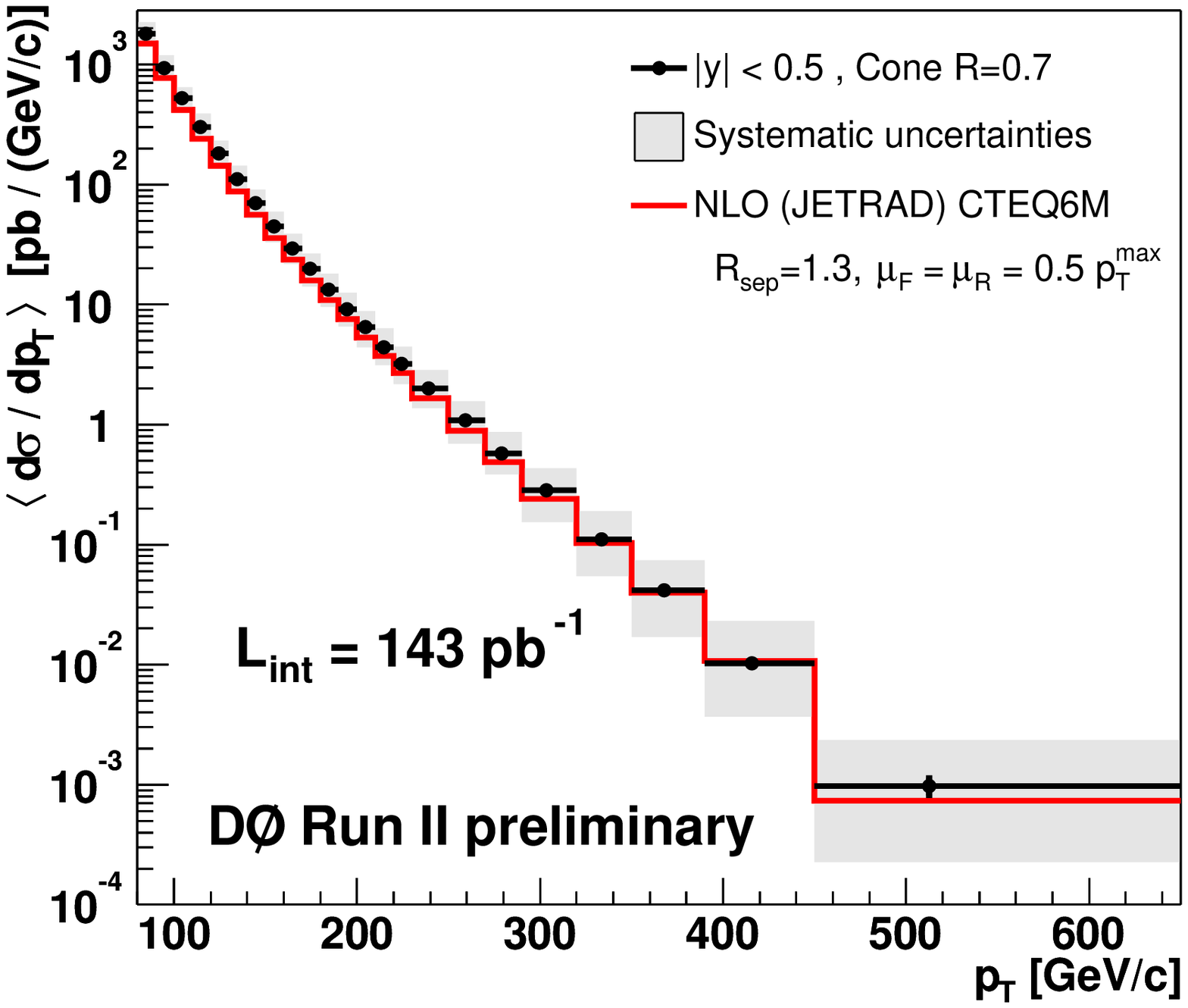}
      \hspace{0.3cm}
      \includegraphics[width=6.7cm,height=6.5cm]{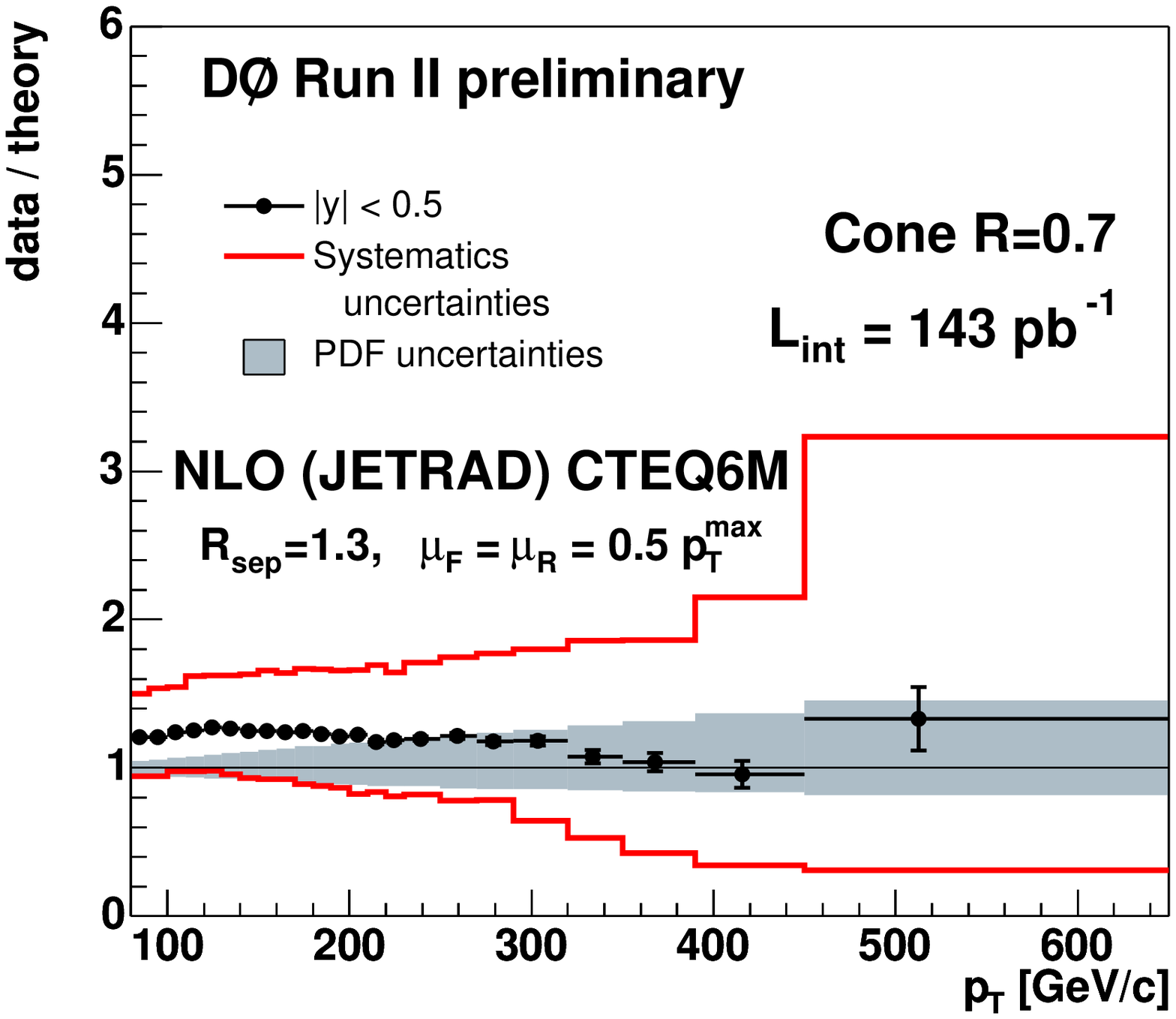}
      }
    }
\caption{Recent preliminary D0 jet data \cite{run2data}. }
\end{figure}

The LHC program for
the discovery of new particles and/or new strong interactions
at high energy relies heavily on the factorization formalism.
So long as it involves heavy states, each prediction of
an extension of the  standard model appears in the
hard scattering function of the factorized cross section.
The parton distributions are the same as those
observed in DIS at lower energy, now extrapolated
by evolution.

\subsection{Factorization proofs and matrix elements}

In the previous section we have made extensive use of
factorization for the DIS structure functions, and now we
have seen how they generalize to hadronic collisions.
Let us come back briefly to the proof of the forms
(\ref{disfact}) for the former and (\ref{hhfact}) for the latter.

For deep-inelastic scattering, there are a  number
of ways to provide a  proof.  One way is based on
the light-cone expansion of Eq.\ (\ref{JJlcexp}), and
is most conveniently formulated in terms of moments.

Alternatively, we may begin wiith the
pinch surfaces and hence physical pictures for
the process $eN\rightarrow e + X$, where we
assume a large momentum transfer, $Q$.  These are shown
in Fig.\ \ref{DISps}, and generally involve jets  in the final state,
in addition to collinear fragments of the incoming nucleon.
The figure shows a single parton, labeled $i$ combining
with the off-shell photon of labeled $q$ to produce a final state.
Physical pictures with more collinear partons are possible.
As in the discussion of jet production in $\rm e^+e^-$ annihilation
above,
at leading power in $Q$, they can be only
unphysically-polarized gluons (in covariant gauges).
\begin{figure}[ht]
\centerline{\epsffile{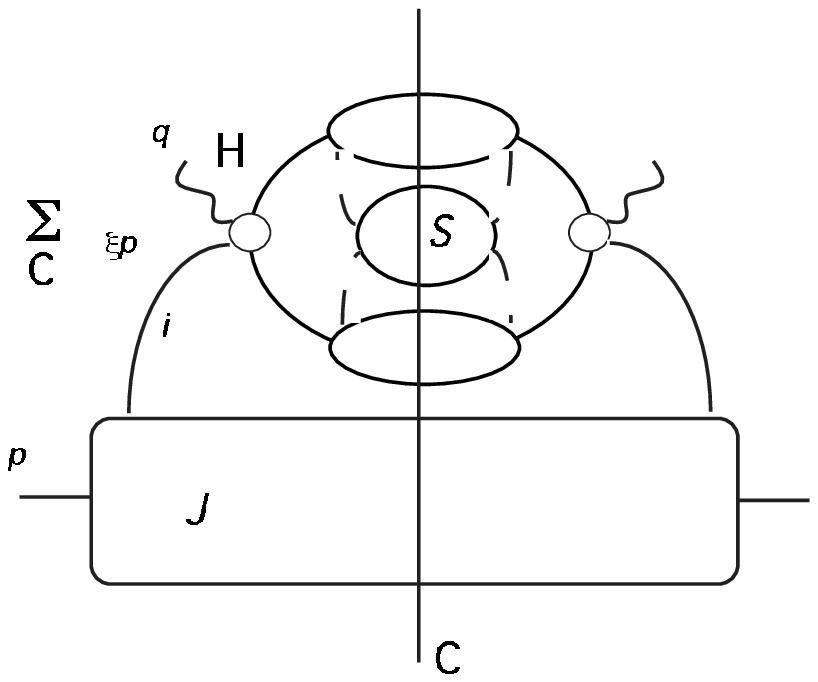}}
\caption{Pinch surface for DIS.}
\label{DISps}
\end{figure}
It is worth noting that the requirement of a
physical  picture implies that the momentum fraction
of the parton that initiates the hard scattering obeys
$1 > \xi > x$, for much the same reasons as in our
low-order examples above; all the partons in a jet
or hadron must be moving in the same direction.
For the inclusive DIS cross section, we  sum over
all final states, and information on the  time development
of the final state jets  is  lost, just as in
the total $\rm e^+e^-$
annihilation cross section.  For DIS, however, the
initial-state evolution that provides the ``active" parton
(the one that initiates the hard scattering) remains.

The DIS argument regarding the sum over states
has the same content as the optical theorem for DIS:
the total $\gamma^*$ p cross section is related to the $\gamma^*$p
forward
scattering amplitude.  In terms of the  hadronic tensor of Eq.\
(\ref{Wtensors}),
this can be written as
\bea
W_{\m\n} = 2\; {\rm Im}\, T_{\m\n}\, ,
\label{disoptical}
\eea
where the tensor $T_{\m\n}$ is the expectation of the
time-ordered product of currents,
\bea
T_{\m\n} = {i\over 8\pi} \int d^4x\, e^{i q \cdot x}  < N (p) |\; T\;
J_\m (x)
J_y (0)\; | N (p)>\, .
\label{Tmunudef}
\eea
Because of the optical theorem, the long-distance behavior of the
hadronic tensor
is specified by the pinch surfaces of this matrix element,
which can be described as the amplitude for forward  Compton scattering
of a nucleon by an off-shell photon.  The pinch surfaces
in this case are much simpler than those for
individual final states.  The requirement of a physical picture shows
that
no jets are possible aside from the ``forward" jet of fragment
partons from the incoming
nucleon, which reform after the hard scattering into an
outgoing nucleon of the same total momentum\footnote{A related physical
off-diagonal amplitude \cite{dvcs}, $\gamma^*+p\rightarrow \gamma+p$
called deeply-virtual Compton scattering,
offers complementary information on hadronic structure, and may also
be factorized \cite{dvcsfact}.}.  The
hard scattering is reduced to a point, because, as
in the $\rm e^+e^-$ case, on-shell particles that travel
a finite distance from the hard scattering can never recombine.
\begin{figure}[ht]
\centerline{\epsffile{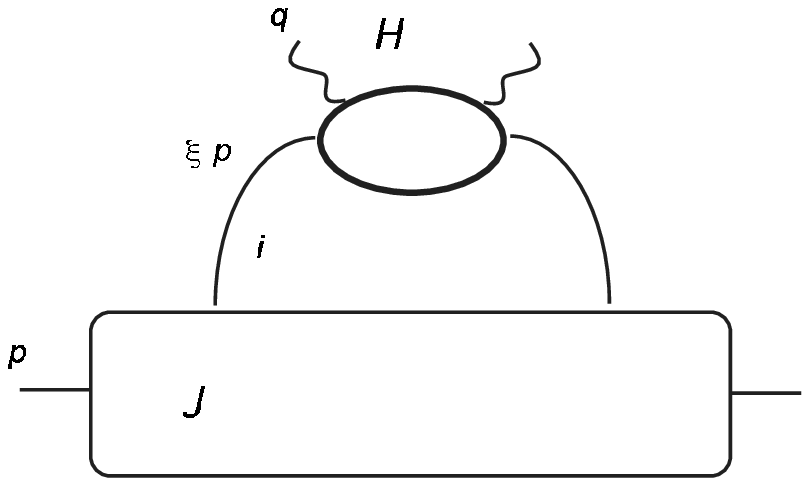}}
\caption{Pinch surface for forward Compton scattering.}
\label{Cptps}
\end{figure}
At the pinch surface, the hard-scattering function $H$
depends only on the non-vanishing longitudinal component $\xi p$
of the active parton's momentum.
Once we have reached this stage, the remainder of the
factorization argument is a systematic disentangling of
the upper and lower parts of Fig.\ \ref{Cptps}, in terms
of color, spin and momentum.
A hierarchy of subtractions \cite{cssrv}
effects this separation.

The parton distributions factorized in this mannner take the
form of the expectation values of operators relatively on the light
cone.
To specify these matrix elements, it is convenient to choose
the momentum of incoming nucleon $h$ in the plus light-cone direction,
and to define an opposite-moving light-like vector, $n^\mu=\delta_{\m
-}$.
The parton distribution is then
\bea
   \f_{q/h} (\x, \m^2) = {1\over 2}
\sum_\s \int^\infty_{-\infty}  {d y^-
\over
2\p} e^{-i \x p^+ y^- }
\,
<  h(p, \s) \mid \bar{q} (0^+, y^-, {\bf 0}_\perp)
\, {1\over 2}n \cdot \gamma\;
   q (0) \mid  h(p, \s) >\, ,
\label{phiqazero}
\eea
where the average over spins for an unpolarized
cross section is exhibited.  This matrix element can
be defined in $n\cdot A=0=A^+$ gauge.
Choosing the incoming hadron as a quark,
and computing (\ref{phiqazero}) at order
$\as$, we find precisely the $\overline{\rm MS}$
quark distribution discussed in Sec.\ 4 above.
The formalism can readily be extended to
covariant gauges, and  the matrix element
made gauge invariant, by connecting the  quark
operators with ordered exponentials (also known
as nonabelian phases, gauge links and Wilson lines),
\bea
{\bar q}(y^-)\; n\cdot \gamma\; q(0)
&\rightarrow&
\bar{q} (y^-) P \exp \left[ i g \int_0^{y-} d \l\;
n\cdot A (\l n^\m) \right] n\cdot \gamma\; q (0)\, .
\label{oexps}
\eea
These and related forms of parton distributions,
including their generalizations to transverse momentum
distributions, were discussed extensively
in Ref.\ \cite{CoSo82}.

An extensive discussion of proofs of perturbative
factorization for semi-inclusive hard scattering
cross sections in hadron hadron scattering \cite{factproofs} may be
found in \cite{cssrv},
with additional comments in Ref.\ \cite{ste95}.
Here, I will only  recall a few of the heuristic arguments.

The essential challenge when there are two hadrons in the initial
state is to show the universality of the parton
distribution functions between DIS and hadron-hadron scattering.
Consider a Drell-Yan cross section, in which a  quark from
one hadron annihilates with an antiquark from the other
to form a virtual photon, W or Z (generic mass $Q$).   We'll think of a
range in kinematics
where the quark and antiquark carry a large fraction $x$
of their parent hadron's momentum, so that there is
no confusion about which of the pair comes from which hadron.

It is natural to wonder whether, with two hadrons in the initial state
the color fields of each hadron might permeate the other
prior to the annihilation, modifying the colors, or even
transverse momenta, of the  annihilating
pair in such a way as to increase or decrease their
amplitude for the annihilation.
A proof of factorization must show that such effects decrease
as a power of the pair invariant mass, $Q$.  The technical
arguments rely on the formalism we have developed above
in Sec.\ 3.   The  summation over final states that defines
the semi-inclusive cross section  results in an
expression for which the only pinch surfaces that
remain logarithmic in their power counting after the sum over final
states
are in one-to-one correspondence with those found in
the inclusive DIS cross section, separately for
the two hadrons.   The hadron-hadron cross section is then
factorizable in the same manner as deep-inelastic scattering.

The physical reason for this rather simple result is
not far to seek.  The Lorentz contracted fields of incident particles
do not overlap
until the moment of the scattering  \cite{bas83}.  As a  result,
initial-state interactions that couple the
two hadrons disappear at high enough energies.
All remaining initial-state interactions are internal
to the hadrons, and are specified by the  same parton
densities as in deep-inelastic scattering.
It is a matter of relativistic causality: as the
relative velocities of the two hadrons approaches the
speed of light, they are unable to exchange signals of any kind.
Correspondingly, the annihilation process itself (or other
hard-scattering
reaction) must also occur on a very short time scale,
and final-state interactions, after the hard scattering,
are simply too late to affect large momentum
transfer and/or creation of the heavy particle(s).

   \subsection{Resummations: a closer look at the final state}

   These lectures have concentrated on applications of
   asymptotic freedom, infrared safety and factorization
   in semi-inclusive cross sections with only a single
   hard scale.  A closer look at the final state, in terms of
   shape functions, generally introduces another  scale,
   as noted in Sec.\ \ref{jetshapesec}.  The second scale
   may be perturbative, but ratios of the two perturbative
   scales may be large at any fixed order in perturbation
   theory.  For example, in the limit of low jet masses in
   Eq.\ (\ref{tMreln}) (thrust goes to 1) or of low
   angularities $e_a$ in Eq.\ (\ref{angdef}),
   these corrections build up as $(1/M_{Ji}^2)\,
\as^n\ln^{2n-1}(M_{Ji}/Q)$
   and $(1/e_a)\as^n\ln^{2n-1}e_a$, respectively.  Note
   the overall additional power singularity at zero jet
   mass in each case.

   In cases like jet masses and angularities, we can control
   these large corrections to all orders in perturbation theory,
   a process known as resummation.  Resummations, like
   evolution in DIS, can usually be derived from a factorization
   formula, as in Eq.\ (\ref{separation}) for evolution.
   We can illustrate this process for the angularities (\ref{angdef}).
   Details are given in Ref.\ \cite{ber03a}, but the basic
   observations are simple enough.  In the limit of small
   $e_a$, the final state consists of two well-collimated jets,
   accompanied by very soft wide-angle radiation.
   The soft radiation cannot resolve the internal structure of
   the jets, and is equivalent to the soft radiation emitted
   by a  pair of point-like recoil-less sources in the  fundamental
   representation of the gauge group (quark and antiquark
   representations).  In this limit, the
   cross section factorizes as shown in Fig.\ \ref{2jetea}.
   \begin{figure}[h]
\hbox{\hskip 2 cm{\epsfxsize=12cm \epsffile{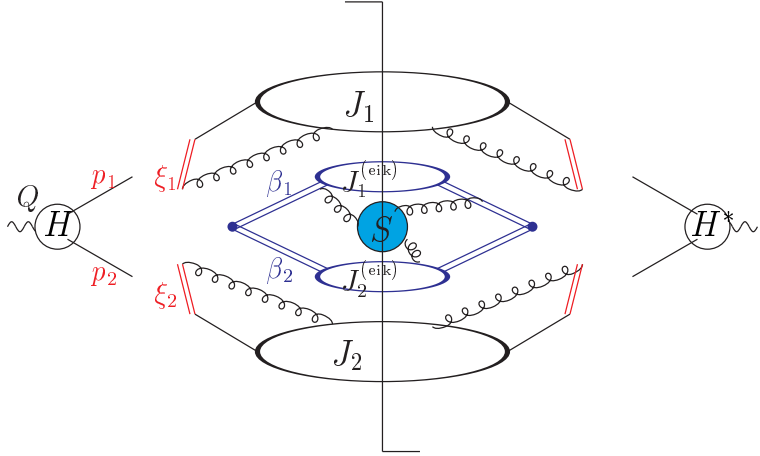}}}
   \caption{Factorization near the two-jet limit.}
   \label{2jetea}
   \end{figure}
Here each of the  individual factors can be given
   a field-theoretic interpretation, an example being the
   functions associated with  the jets,
   \bea
\bar{J}_q^\mu (p\cdot \xi,e_a,\mu)
&=&
     \frac{(2\pi)^6}{N_c} \; \sum\limits_{N}
{\rm Tr} \Bigg[\gamma^\mu
\left<0 \left|
\Phi^{(\rm q)}_{\xi}{}^\dagger(0,-\infty;0) q(0) \right| N \right>
    \left< N\left| \bar{q}(0) \Phi^{(\rm q)}_{\xi}(0,-\infty;0)
\right| 0 \right>\Bigg ] \nonumber \\
& &\, \hspace{10mm} \times \, \delta(e_a-
e_a(N,a))\,\delta^3\left(p_{J} - p(N)\right)\, .
\label{jetdef}
\eea
As in the case of the matrix element that defines the quark
distribution
   (\ref{phiqazero}) with (\ref{oexps}),
   we have introduced an ordered exponential
   in the direction of the vector $\xi$ to make the
   jet function gauge-invariant.  We also need to
   introduce a factorization scale, $\m$, to separate
   the hard, jet and soft functions.
   Both the scale $\mu$ and the vector $\xi^\n$
are artifacts of the factorization; neither appears
in the cross section itself, but the factorization
is valid up to positive powers of $e_a$, which
is small.  The factorized cross section therefore
obeys two equations, analogous to (\ref{physzero}),
\bea
\mu {d\sigma\over d\mu} &=& 0
\nonumber\\
\xi^\alpha {d\sigma\over  d\xi^\alpha} &=& 0\, .
\label{consistency}
\eea
The solutions to these equations are a generalization
of the evolution of parton distributions, and as in
that case, they are  most simply presented in
a transform space, in this case a Laplace transform
with respect to the angularities.  The  limit of small $e_a$
   corresponds directly to large values of the
   transform variable $\n$, dependence on which exponentiates,
   \bea
\sigma(\nu,Q,a)
&=&
\int_0 de_a\, {\rm e}^{-\nu e_a}\;
\sigma(e_a,Q,a)= {\rm e}^{E(\nu,Q,a)}\, ,
\label{sigmalaplace}
\eea
   where the exponent $E(\nu)$ is given by
   \bea
\hspace{-80mm}
E(\nu,Q,a) &=&
    2\, \int\limits_0^1 \frac{d u}{u} \Bigg[ \,
        \int\limits_{u^2 Q^2}^{u Q^2} \frac{d p_T^2}{p_T^2}
A_q\left(\as(p_T)\right)
        \left( {\rm e}^{- u^{1-a} \nu \left(p_T/Q\right)^{a} }-1 \right)
\nonumber \\
        & & \hspace{5mm}      + \frac{1}{2} B_q\left(\as(\sqrt{u}
Q)\right) \left( {\rm e}^{-u
\left(\nu/2\right)^{2/(2-a)} } -1 \right)
        \Bigg] \, .
\label{Eofnu}
\eea
In the exponent, the presence of
two integrals reflects the two consistency equations
(\ref{consistency}), and the
functions
$A_q$, $B_q$ are infrared-safe anomalous dimensions.
In particular
$A=(\as/\pi)\, C_F + \dots $ also appears in $P_{qq}(x,\as)=
A(\as)/(1-x) + \dots$,
and is now known to three loops \cite{as3split}.
For quark jets, $B=(\as/\pi)(3C_F/2) + \dots$ is somewhat more
process dependent.
Expressions like this, with a double integral
over the running coupling, were first derived in studies
of the transverse momentum distribution for the  Drell-Yan
cross section, especially as developed by  Collins and Soper \cite{bbj}
and Sen \cite{sensud}.

The inversion of the transform leads to distributions of the sort
shown in Fig.\ \ref{heavyjet} for the mass of the ``heavy" jet in
$\rm e^+e^-$ annihilation, and for the distribution of
Z transverse momenta at the Tevatron in Fig.\ \ref{Zqt}.
\begin{figure}[h]
\centerline{\epsfxsize=10cm \epsffile{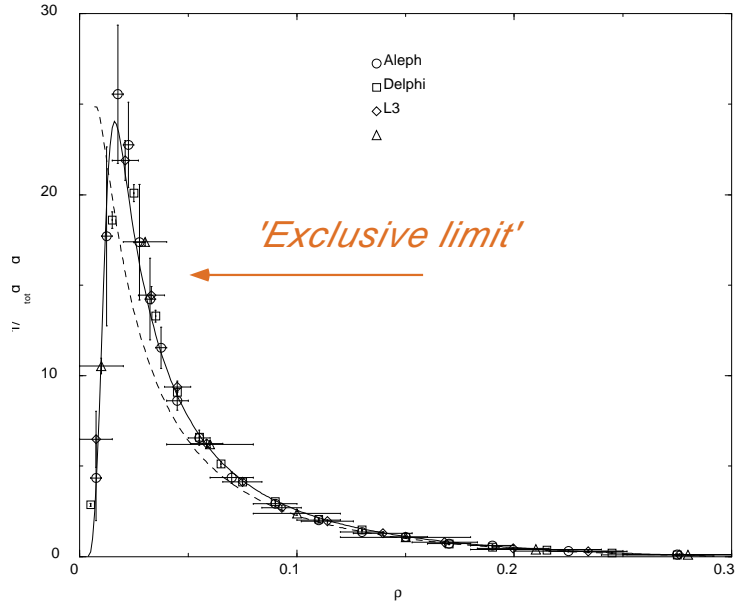}}
\caption{Heavy jet distribution.  From \cite{kor00}.}
\label{heavyjet}
\end{figure}
\begin{figure}[h]
\hbox{\hskip 2 cm {\epsfxsize=12cm \epsffile{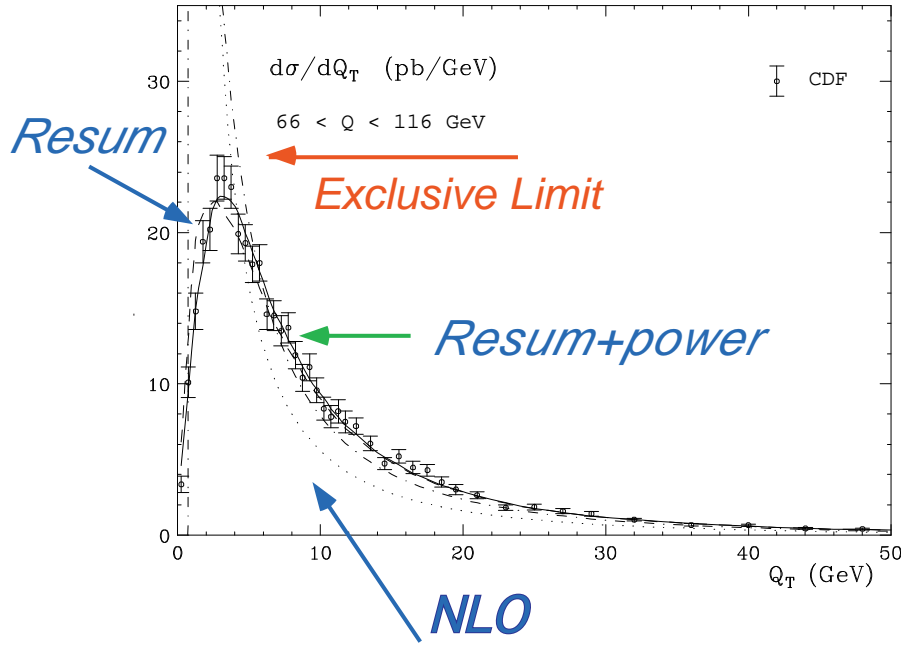}}}
\caption{Transverse momentum distribution for the Z at
the Tevatron.  From \cite{kul02}.}
\label{Zqt}
\end{figure}

In both figures, the cross section has a peak very near
the kinematic configuration of elastic scattering:
for $\rm e^+e^-$,  $e_a=0$, corresponding to a quark-pair
final state, and $Q_T=0$, corresponding to the
production of a Z in Born approximation.  In both cases,
the effect of gluonic radiation is to turn a delta function
distribution into one that rises to a peak near the
elastic limit, and then slowly falls off away from that limit.
The tail of the distribution away from the peak is given by
low-order perturbation theory, and the position of the
peak can be calculated perturbatively as well.
Nevertheless, the bulk of the inclusive cross section
is concentrated near the peak, and in this region,
nonperturbative corrections are suppressed only
by powers off  $1/(e_aQ)$, rather than $1/Q$ \cite{shaperv}.  This
effect can be inferred directly from the Eq.\ (\ref{Eofnu})
\cite{kor99}.
A sketch of the transition from resummed
perturbation theory to nonperturbative power corrections
can be found in Ref.\ \cite{gsismd04}.

\subsection{Higher orders}

Predictions based on factorized
cross sections in perturbative QCD require both
hard-scattering functions and parton distributions.
It is an understatement that making progress on both
of these topics is nontrivial in almost every case of practical
interest.
The past few years, however, have seen important progress,
through a growing mastery over higher order corrections, and
through the improved determination
of parton distributions from a growing set of data.  Certainly, the
coming of the LHC has helped drive much of this
work, but their own internal logic, and the special challenges
these projects pose, have surely provided
strong motivations as well.

\subsubsection{Toward a two-loop phenomenology}

In effect, the transition from leading order to
next-to-leading order accuracy at the level of
jet cross sections took about a decade, and a similar
time scale will have been necessary to realize,
for example, full NNLO jet cross sections in time
for the LHC.  Given the experimental value of
$\as(m_{\rm Z})\sim 1.2$ \cite{as04}, hard-scattering cross sections at
the
order of $\as^2(M_{\rm Z})$ have a nominal accuracy in the per cent
range.  For some semi-inclusive cross sections at momentum transfers in
the hundreds of GeV,
this is also the scale of many nonperturbative corrections, including
those
associated with uncertainties in the parton distribution functions (see
below).
Percent accuracy is  probably attainable for benchmark cross sections,
as a foundation from which to understand the complex final states
associated with the decay of new, heavy particles, and the  QCD
backgrounds to them.

The development of a NNLO phenomenology is following
the pattern of NLO.  For some time, fully inclusive NNLO cross sections
for $\rm e^+e^-$
annihilation, DIS, and Drell-Yan annihilation
(including Higgs production) have been available \cite{inclusivennlo}.
More complex is the combination of final states in $\rm e^+e^-$ jet
cross sections,
and even more so the factorization necessary for
jet cross sections in hadron-hadron scattering.
Even at NLO the state of the art is 4-jet cross sections for $\rm
e^+e^-$,
and up to 3 jets at hadron colliders \cite{nlojets}.

Within the last few years, the elements necessary for factorized NNLO
cross sections are beginning to be assembled in earnest.  Progress has
been
made in NNLO $\rm e^+e^-$ jet cross sections \cite{epem2jnnlo},
while the amplitude at ${\cal O}(\as^3)$ has become available for
three jets.  The past few years have also  seen the first NNLO two-loop
S-matrix elements for the $2\to 2$ scattering \cite{2to2nnlo}, an
essential step
toward jet cross sections in hadron-hadron scattering.
Important progress is also ongoing toward organizing the
singular integrals over phase space \cite{phasespacennlo}.
Equally significant
is the calculation of NNLO splitting functions, the fruit of a
decade-long
program \cite{as3split}.

Many of these acheivements have required developing new
methods in the organization of large numbers of
Feynman diagrams \cite{dixont95} often involving
astronomical numbers of terms.   Hope springs eternal that
new mathematical insights \cite{twistor} may simplify the road ahead,
at least for the calculation of on-shell amplitudes.

\subsubsection{Parton distributions: global fits and uncertainties}

The determination of parton distributions is a bootstrap process.
As described in Sec.\ 3, they must be abstracted from the
comparison of sets of data to factorized cross sections,
with hard-scattering functions computed up to a given
order.   For the past ten years, the calculation has been at the level
of  NLO.
Any set of distributions determined this way is
limited not only by the accuracy of the calculations,
but also  by the sensitivity of the  data to given distributions.
For example, in DIS, photons or weak vector bosons
couple to the quarks at zeroth order, while the
gluons get into the act only at order $\as$ in the
hard scattering.  Naturally, we might expect
DIS to be less informative about gluons than  about quarks,
and we look for other processes, like jet production, which
can be more directly sensitive to  the gluon distribution.
Other cases where new data have provided new
and sometimes surprising insights concern the ratios
of  d-quark  and u-quark distributions from
W$^\pm$ asymmetry measurements at the
Tevatron collider \cite{dycdf}, and of the $\bar{\rm d}$
and $\bar{\rm u}$ from fixed-target Drell-Yan data \cite{e866}.

In any case, once a model set of PDFs has been produced,
it provides predictions for other factorizable cross sections.
As new data become available, it may require us to
modify existing models to incorporate new features.
This feedback procedure has evolved into the
approach of ``global" fits,  which try to  use the best
available data from a variety of cross sections.
(Naturally, these are cross sections for which
it is anticipated that no new physics signals
are confusing things!)

The leading global fits have been provided by the
CTEQ \cite{cteq6} and MRST \cite{mrst} collaborations.
Sets of distributions are given in terms of the parameters
specified in recent work for each partons by
   \bea
   xf(x,Q_0) = A_0 x^{A_1} (1-x)^{A_2} {\rm e}^{A_3x} (1+ e^{A_4} 
x)^{A_5}
   \eea
   for each of the CTEQ partons, and
   \bea
   xq(x,Q_0) &=& A(1-x)^\eta(1+\epsilon x^{0.5} + \gamma x) x^\delta
   \nonumber\\
    xg(x,Q_0) &=&
A_g(1-x)^{\eta_g}(1+\epsilon_gx^{0.5}+\gamma_gx)x^{\delta_g}
    - A_-(1-x)^{\eta_-}x^{-\delta_-}
   \eea
for the MRST quarks and the gluon, respectively.
Figure \ref{pdfcompare} shows a broad agreement between recent sets,
but also the difference in some details.
\begin{figure}[h]
\centerline{\epsfxsize=9cm \epsffile{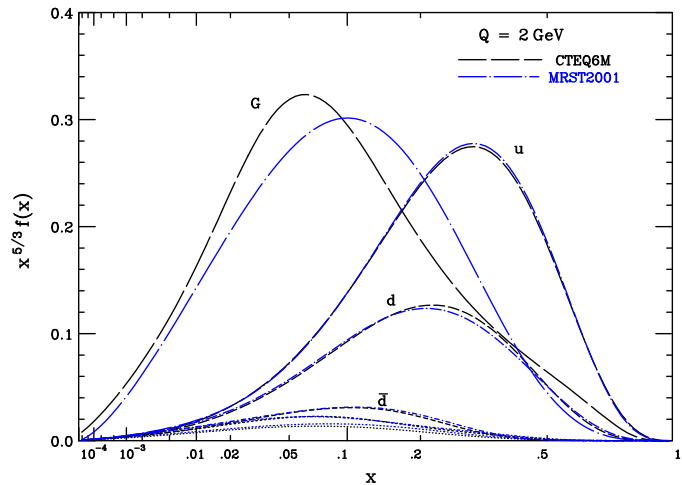}}
\caption{Recent parton distribution fits.}
\label{pdfcompare}
\end{figure}
The parameters in these functions are fitted to the data,
but have little physical significance in themselves,
aside from certain built-in structure, such as the vanishing
of the  distributions at $x=1$.  An  important development of
the past few years is the drive to quantify uncertainties
within the sets \cite{cteq6},
by analyzing the response of the
fits as the parameters are varied.
It is now routine to see uncertainties based
on these analyses quoted by experiments alongside
their more familiar systematic and statistical errors.

\section{Closing Comments}

In these lectures, we have only scratched the surface
of the many applications of perturbative QCD, let
alone the many and varied explorations of its
nonperturbative structure.   From the former, evolution
at small $x$, and its relation to nuclear collisions,
the analysis of heavy quark decays, polarized scattering
effective field theory approaches, and developments
in multi-loop calculations are just a few very active areas.
I hope, however, that these lectures may
provide a useful introductory perspective  to some of the ideas
underlying this vast area of research.

\subsection*{Acknowledgments}

I would like to thank the organizers of TASI 2004,
John Terning, Carlos Wagner, and Dieter Zeppenfeld,
for the invitation to lecture.  I am also grateful to K.T.\ Mahanthappa,
and the TASI school for supporting my visit.   I feel lucky to
have had the opportunity to interact with the TASI 2004 students, and I
thank them
for the interest they showed in the subjects discussed here.
This work was supported in part by the National Science
Foundation, grants PHY-0098527 and PHY-0354776.

\end{document}